\documentclass[aps,pre,reprint,groupedaddress]{revtex4-1}

\usepackage{amsmath} 
\usepackage{graphicx}
\usepackage{bm}
\usepackage{booktabs}

\begin{document}


\title{Optimal short-term memory before the edge of chaos in driven random recurrent networks}


\author{Taichi Haruna}
\email[]{tharuna@lab.twcu.ac.jp}
\affiliation{Department of Information and Sciences, School of Arts and Sciences, Tokyo Woman's Christian University, 2-6-1 Zempukuji, Suginami-ku, Tokyo 167-8585, Japan}

\author{Kohei Nakajima}
\email[]{k\_nakajima@mech.t.u-tokyo.ac.jp}
\affiliation{Graduate School of Information Science and Technology, University of Tokyo, Bunkyo-ku, Tokyo 113-8656, Japan}


\date{\today}

\begin{abstract}
The ability of discrete-time nonlinear recurrent neural networks to store time-varying small input signals is investigated by mean-field theory. The combination of a small input strength and mean-field assumptions makes it possible to derive an approximate expression for the conditional probability density of the state of a neuron given a past input signal. From this conditional probability density, we can analytically calculate short-term memory measures, such as memory capacity, mutual information, and Fisher information, and determine the relationships among these measures, which have not been clarified to date to the best of our knowledge. We show that the network contribution of these short-term memory measures peaks before the edge of chaos, where the dynamics of input-driven networks is stable but corresponding systems without input signals are unstable. 
\end{abstract}


\maketitle

\section{Introduction}
\label{sec:intro}
Natural and artificial high-dimensional nonlinear dynamical systems can be used as resources for real-time computing. By nonlinearly mapping time-varying input signals into a high-dimensional space, the signals can be learned in a supervised manner if the dynamical systems have enough ability to store the signals in their present state and separate different signals~\cite{Jaeger2001,Maass2002}. A high computational performance can be achieved by tuning only the weights of linear connections to the output layer while keeping the parameters of the dynamical systems fixed~\cite{Jaeger2004,Verstraeten2007,Lukosevicius2009, Pathak2018}. Such dynamical systems called reservoirs can be artificial recurrent neural networks (RNNs) or physical systems, such as optical media~\cite{Larger2012,Larger2017}, nanoscale magnetization dynamics~\cite{Torrejon2017, Tsunegi2019}, soft materials~\cite{Nakajima2018}, and quantum systems~\cite{Fujii2017}. 

As mentioned above, a requirement for real-time computing is the ability to memorize past input signals. Such short-term memory of dynamical systems has been studied extensively by assessing a quantity called memory capacity~\cite{Jaeger2002,Dambre2012}. For input-driven RNNs, it has been suggested that the part of memory capacity representing indirect memory through network takes a maximum value near the edge of chaos, namely, near the critical boundary between the stable and unstable dynamical regimes~\cite{Bertschinger2004,Boedecker2012}. Near criticality, different inputs are expected to lead to different states while suppressing the influence of the initial conditions. Hence, it seems reasonable for a dynamical system to be near the critical point for optimal memory capacity. However, it has also been pointed out that the dependence on network parameters is not straightforward based on a systematic numerical simulation~\cite{Farkas2016}. 

For linear RNNs, detailed analytic studies of memory capacity can be performed for both discrete-time~\cite{White2004,Rodan2011} and continuous-time systems~\cite{Hermans2010}. The ability to predict future inputs, which is complementary to memory capacity, has also been studied in linear systems with correlated input signals~\cite{Marzen2017}. However, the memory capacity of nonlinear RNNs is difficult to study by analytical methods~\cite{Ganguli2008}. Recently, Schuecker et al.~\cite{Schuecker2018} successfully derived an analytical expression for memory capacity for continuous-time nonlinear RNNs~\cite{Sompolinsky1988} in which each neuron is driven by independent input signals following a white-noise Gaussian process. Toyoizumi and Abbott~\cite{Toyoizumi2011} analytically calculated the signal-to-noise ratio, which is equivalent to the inverse of memory capacity at the limit of zero input strength, for discrete-time nonlinear RNNs driven by a common time-varying input signal. 

In this paper, we analytically investigate the memory capacity of discrete-time nonlinear RNNs called echo state networks (ESNs)~\cite{Jaeger2001} by a mean-field theory when the strength of input signals is small but non-zero. The main idea of our approach is that the conditional probability density of the present state of a neuron given a past input signal can be approximately calculated from a functional derivative with respect to past input signals under the assumption of a small input strength. Once we obtain this conditional probability density, it is straightforward to derive the memory capacity and other alternative memory measures, such as mutual information and Fisher information~\cite{Ganguli2008}. We show that all three measures of short-term memory through network behave similarly and take a maximum value before the edge of chaos, where the dynamics is stable in the presence of input signals but unstable in the absence of input signals. We also discuss the breakdown of the mean-field theory for calculating memory measures in the ordered regime and show that the linear approximation provides good predictions. 

\section{Results}
\label{sec:results}

\subsection{Echo State Networks}
\label{sec:esn}
We consider ESNs consisting of $N$ artificial neurons. The state of neuron $i$ at discrete time step $t$ is denoted $x_i(t)$. We assume that the time evolution of state $x_i(t)$ is governed by 
\begin{equation}
x_i(t+1)=f(a_i(t)), 
\label{eq:1}
\end{equation}
where $f$ is an activation function. $a_i(t)$ is the activation potential of neuron $i$ at time step $t$ given by 
\begin{equation}
a_i(t)=\sum_{j=1}^N w_{ij} x_j(t) + u_i s(t), 
\label{eq:2}
\end{equation}
where $s(t)$ is a time-dependent input signal, $w_{ij}$ is a time-independent weight of the connection from neuron $j$ to neuron $i$, and $u_i$ is a time-independent weight representing the strength of the coupling from the input signal to neuron $i$. We use the matrix and vector notations $W:=\left( w_{ij} \right)_{1 \leq i,j \leq N}$, $\bm{u}:=(u_1,u_2,\dots,u_N)^\textrm{T}$, and $\bm{x}(t):=(x_1(t),x_2(t),\dots,x_N(t))^\textrm{T}$. 

In the following analytical calculations and numerical simulations, the activation function is assumed to be a sigmoid function satisfying $\lim_{a \to \pm \infty} f(a)=\pm 1$, $f(-a)=-a$ and $f'(0)=1$. In particular, we adopt $f(a)=\textrm{erf}(\frac{\sqrt{\pi}}{2}a)$ owing to its analytical tractability. $w_{ij}$ are chosen independently at random from an identical Gaussian distribution with mean zero and variance $g^2/N$, where $g^2>0$ is a control parameter. For simplicity, $u_i$ are assumed to be independent variables taking $\pm 1$ with probability $\frac{1}{2}$. Since our primary concern is the memory capacity of ESNs, we consider an independent and identically distributed Gaussian input signal $s(t)$ with mean zero and variance $s^2$. All numerical results in this paper were obtained in the following way unless otherwise stated. We simulated ESNs with $N=1000$ artificial neurons over $40$ trials. For a single trial, each quantity (for example, stationary variance of $x_i(t)$) was calculated from its values over $10^5$ time steps after discarding the initial $10^4$ time steps. Then, averages were obtained over all artificial neurons and all trials. All infinite sums appearing in the following sections were evaluated by truncation at the $500$-th term. 

\begin{figure}[t]
\centering
\includegraphics[width=8cm]{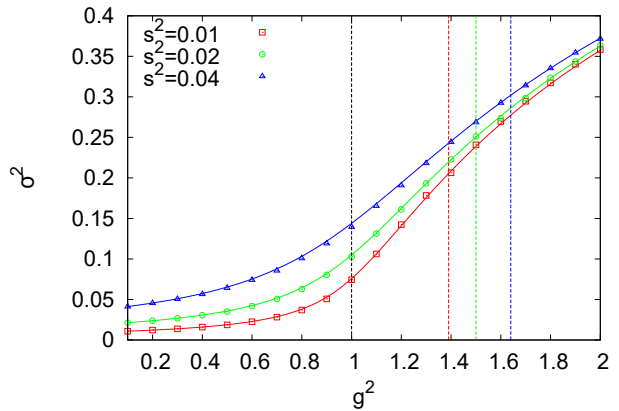}
\caption{
Numerical results (marks) and mean-field predictions (solid lines) for the stationary variance of $x_i(t)$ are shown as functions of $g^2$ for $s^2=0.01$ (red), $s^2=0.02$ (green), and $s^2=0.04$ (blue).  Vertical broken lines indicate the mean-field predictions of the critical value of $g^2$ obtained from Eq.~(\ref{eq:4}) for $s^2=0$ ($g^2=1$, black), $s^2=0.01$ ($g^2 \approx 1.39$, red), $s^2=0.02$ ($g^2 \approx 1.50$, green), and $s^2=0.04$ ($g^2 \approx 1.64$, blue). 
}
\label{fig:1}
\end{figure}

The mean-field theory of ESNs~\cite{Cessac2007,Massar2013,Molgedey1992,Toyoizumi2011} makes it possible to calculate the stationary variance of $x_i(t)$ and the largest Lyapunov exponent in the limit $N \to \infty$. It assumes that $x_i(t)$ are independent and identically distributed random variables. They are also assumed to be independent of $w_{ij}$ and $u_i$. This assumption can be justified in the limit $N \to \infty$ when there is no input signal. Since $f$ is odd, we can self-consistently assume that the mean of $x_i(t)$ is equal to zero. Let $\sigma^2(t)=\langle x_i(t)^2 \rangle$ be the variance of $x_i(t)$, where $\langle \cdots \rangle$ indicates the average over trials with the same $w_{ij}$ and $u_i$ but possibly different realizations of the input signal $s(t)$ and initial conditions. By the central limit theorem, $a_i(t)$ follows a Gaussian distribution with mean zero and variance $g^2 \sigma^2(t)+s^2+ O(N^{-\frac{1}{2}})$, where we use $u_i^2=1$. Neglecting the $O(N^{-\frac{1}{2}})$ term, the variance of $a_i(t)$ does not depend on specific realizations of $w_{ij}$~\cite{Toyoizumi2011}. In the following, we omit quantities that approach $0$ as $N \to \infty$ unless otherwise stated. Consequently, $\sigma^2(t)$ follows the following recurrence equation~\cite{Massar2013}: 
\begin{align}
\sigma^2(t+1)
&= \int_{-\infty}^\infty da \, f^2(a) \frac{\exp\left( - \frac{a^2}{2\Sigma^2(t)} \right)}{\sqrt{2\pi \Sigma^2(t)}} \nonumber\\
&= -1 + \frac{4}{\pi} \arctan \left( \sqrt{1+\pi \Sigma^2(t)} \right),
\label{eq:3}
\end{align}
where $\Sigma^2(t)=g^2 \sigma^2(t)+s^2$. By numerically solving Eq.~(\ref{eq:3}) with $\sigma^2(t+1)=\sigma^2(t)=\sigma^2$, we can obtain the mean-field prediction of the stationary variance of $x_i(t)$. We write $\Sigma^2=g^2 \sigma^2+s^2$. These values of $\sigma^2$ and $\Sigma^2$ have been used for plotting theoretical results later in the paper. 

The largest Lyapunov exponent derived from the mean-field theory is~\cite{Massar2013,Molgedey1992,Toyoizumi2011} 
\begin{align}
\lambda
&= \frac{1}{2} \log \left( g^2 \int_{-\infty}^\infty da \, f'^2(a) \frac{\exp\left( - \frac{a^2}{2\Sigma^2} \right)}{\sqrt{2\pi \Sigma^2}} \right) \nonumber\\
&= \frac{1}{2} \log \frac{g^2}{\sqrt{1+\pi \Sigma^2}}. 
\label{eq:4}
\end{align}
Here, $\lambda>0$ indicates that a small perturbation to a state of the system leads to exponential growth, while $\lambda<0$ implies that the perturbation eventually becomes undetectable. The dynamics is called chaotic or unstable in the former case and called ordered or stable in the latter. When an input signal is absent, the boundary between chaos and stability $\lambda=0$ corresponds to $g^2=1$. The presence of input signals shifts the boundary towards the chaotic side~\cite{Massar2013,Molgedey1992}. 

In Fig.~\ref{fig:1}, we confirm that the mean-field prediction of the stationary variance of $x_i(t)$ agrees well with the result obtained by numerical simulations. We note that the difference is negligible, even for the input-driven regime $g^2 \ll 1$ where the mean-field assumption is expected to be violated. This will be explained when we discuss the breakdown of the mean-field theory in the calculation of memory capacity. 

\subsection{Memory Capacity}
\label{sec:mc}
The memory capacity of an ESN is defined as the quality of the optimal linear estimator of the past input $s(t-n)$ using the present state of neurons $x_i(t)$. Following previous work~\cite{Schuecker2018,Toyoizumi2011}, we assume a sparse readout, namely, there are $K=O(1)$ readout neurons $1 \leq i \leq K$ and consider linear readout $\hat{s}(t)=\sum_{i=1}^K v_i x_i(t)$. Given time-delay $n$ ($n=1,2,\dots$), the weights $v_i$ are determined by minimizing the mean squared error between $s(t-n)$ and $\hat{s}(t)$ over a sufficiently long time period $T$. The optimal mean squared error as a function of time-delay $n$ is called the memory function and is given by~\cite{Dambre2012,Jaeger2002}
\begin{equation}
M_n = \frac{\sum_{i,j=1}^K \langle s(t-n)x_i(t) \rangle_T d_{ij} \langle x_j(t)s(t-n) \rangle_T }{\langle s(t-n)^2 \rangle_T},
\label{eq:5}
\end{equation}
where $d_{ij}$ is the $(i,j)$-th element of the matrix $C^{-1}$, which is the inverse of the matrix $C=\left( \langle x_i(t)x_j(t) \rangle_T \right)_{1 \leq i,j \leq K}$, and $\langle \cdots \rangle_T$ indicates the time average over the period of length $T$. The memory capacity is defined as the sum of $M_n$ over all time-delays $n$: 
\begin{equation}
M=\sum_{n=1}^\infty M_n. 
\label{eq:6}
\end{equation}
It is known that $M \leq K$ holds~\cite{Jaeger2002,Dambre2012}. 

To compute $M_n$ by the mean-field theory, we replace the time average $\langle \cdots \rangle_T$ for $T \to \infty$ by the average over trials $\langle \cdots \rangle$ in stationary states. Since $\langle x_i(t)x_j(t) \rangle$ for $i \neq j$ vanishes as $N \to \infty$, the contribution of the off-diagonal terms of $C^{-1}$ to Eq.~(\ref{eq:5}) can be neglected for $N \to \infty$ by the sparse readout assumption $K=O(1)$. Thus, in the mean-field calculation, Eq.~(\ref{eq:5}) is just $K$ times $M_n$ for $K=1$. Since 
\begin{equation}
M_n = \frac{\langle x_i(t)s(t-n) \rangle^2}{\sigma^2 s^2} 
\label{eq:7}
\end{equation}
when $K=1$, the task to obtain $M$ reduces to calculating $\langle x_i(t)s(t-n) \rangle$ for $n=1,2,\dots$. In the following, we perform the calculation by assuming that the strength of the input signal is small, namely, $s^2 \ll 1$. 

\begin{figure*}[t]
\centering
\includegraphics[width=16cm]{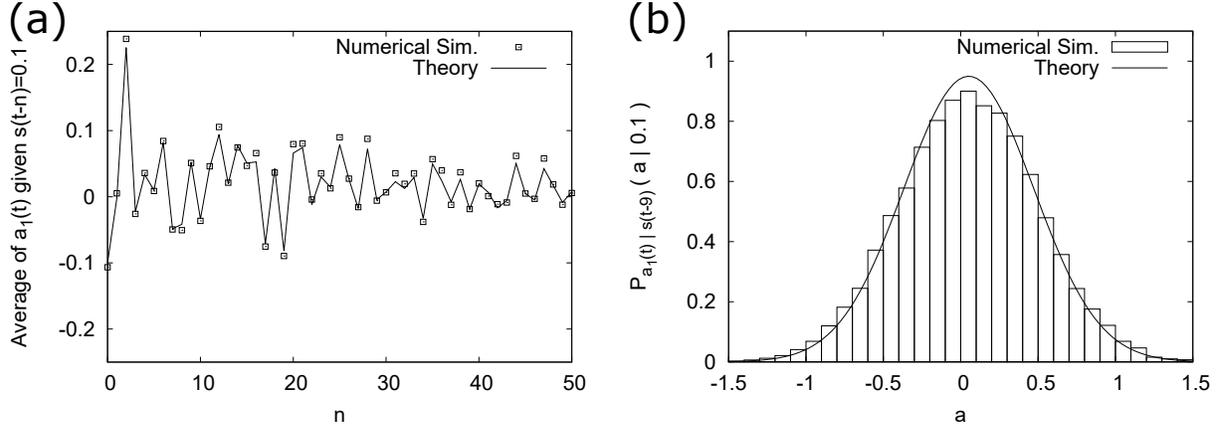}
\caption{
Conditional probability density of the activation potential given a past input. Theoretical predictions and numerical results are compared for (a) the mean of $a_1(t)$ given $s(t-n)=0.1$ and (b) the conditional probability density $P_{a_1(t)|s(t-9)}(a|c)$. We set $s^2=0.01$ and use single specific realizations of $W$ and $\bm{u}$. 
}
\label{fig:2}
\end{figure*}

The main idea to calculate $\langle x_i(t)s(t-n) \rangle$ is that conditioning of $a_i(t)$ on $s(t-n)$ ($n=0,1,2,\dots$) can be regarded as a small perturbation to dependence of $a_i(t)$ on $s(t-n)$, when $s^2 \ll 1$. We assume that $x_i(t)$ are independent and identically distributed and are also independent of $w_{ij}$ and $u_i$ even after conditioning. By the central limit theorem, $a_i(t)$ conditioned on $s(t-n)=c$ follows a Gaussian distribution in the limit of large $N$. Hence, it is sufficient to calculate its mean $\mu_{n,c,i}$ and variance $\Sigma^2_{n,i}$ to determine the conditional probability density $P_{a_i(t)|s(t-n)}(a|c)$. 

First, we calculate the mean of $a_i(t)$ given $s(t-n)=c$. We regard $a_i(t)$ as a functional of stochastic variables $s(t),s(t-1),\dots,s(t-n),\bm{x}(t-n)$. We write $a_i(t)=\mathcal{F}_i \left[\bm{s}_{0:n}(t), \bm{x}(t-n) \right]$, where $\bm{s}_{0:n}(t)=(s(t),s(t-1),\dots,s(t-n))$. We consider a norm defined by the average over trials $\langle \cdots \rangle$ ($\|X\|:=\sqrt{\langle X^2 \rangle}$ for a stochastic variable $X$). Conditioning of $a_i(t)$ on $s(t-n)=c$ corresponds to replacing argument $s(t-n)$ with the constant stochastic variable $c$. If $c^2,s^2 \ll 1$, then $\| c - s(t-n)\|^2=\langle c - s(t-n) \rangle^2=c^2+s^2 \ll 1$. Thus, $a_i(t)$ given $s(t-n)=c$ can be approximated by the following:
\begin{widetext}
\begin{align}
\mathcal{F}_i \left[\bm{s}_{0:n-1}(t),c,\bm{x}(t-n) \right]
&\simeq \mathcal{F}_i \left[\bm{s}_{0:n}(t),\bm{x}(t-n) \right] + \frac{\delta \mathcal{F}_i \left[\bm{s}_{0:n}(t),\bm{x}(t-n) \right] }{\delta s(t-n)} \left( c - s(t-n) \right) \nonumber\\
&= a_i(t) +  \frac{\delta a_i(t)}{\delta s(t-n)} \left( c - s(t-n) \right)
\label{eq:8}
\end{align}
\end{widetext}
By taking the average over trials, we have 
\begin{equation}
\mu_{n,c,i}= \left\langle \frac{\delta a_i(t)}{\delta s(t-n)} \right\rangle c - \left\langle \frac{\delta a_i(t)}{\delta s(t-n)}s(t-n) \right\rangle. 
\label{eq:9}
\end{equation}
By applying the mean-field assumption, we find (Appendix~\ref{sec:a})
\begin{equation}
\left\langle \frac{\delta a_i(t)}{\delta s(t-n)} \right\rangle = \left( V^n \bm{u} \right)_i, 
\label{eq:10}
\end{equation}
where 
\begin{equation}
V=\frac{1}{\sqrt{1+\frac{\pi}{2}\Sigma^2}}W.
\label{eq:11}
\end{equation}
Since $f$ is an odd function, we have (Appendix~\ref{sec:b}) 
\begin{equation}
\left\langle \frac{\delta a_i(t)}{\delta s(t-n)}s(t-n) \right\rangle = 0. 
\label{eq:12}
\end{equation}
From Eqs.~(\ref{eq:10}) and (\ref{eq:12}), the mean-field theory predicts 
\begin{equation}
\mu_{n,c,i}=\left( V^n \bm{u} \right)_i c 
\label{eq:13}
\end{equation}
when $c^2, s^2 \ll 1$. 

Second, the variance of $a_i(t)$ given $s(t-n)=c$ can be obtained as follows. The variance of $a_i(t)$ can be expressed as 
\begin{align}
\Sigma^2 
&= \int_{-\infty}^{\infty} da \, \int_{-\infty}^{\infty} dc \, P_{a_i(t) | s(t-n)}(a|c) P_{s(t-n)}(c) a^2 \nonumber\\
&= \Sigma_{n,i}^2 + \left( V^n \bm{u} \right)_i^2 s^2. 
\label{eq:14}
\end{align}
Thus, we have 
\begin{equation}
\Sigma_{n,i}^2 = \Sigma^2 - \left( V^n \bm{u} \right)_i^2 s^2. 
\label{eq:15}
\end{equation}
Note that the population variance of $\left( V^n \bm{u} \right)_i^2$ takes a nonzero finite value even in the limit of large $N$, as we will see in the linear case (Appendix~\ref{sec:e}) when we discuss the breakdown of the mean-field theory in the ordered regime. This implies that the value of $\Sigma_{n,i}^2$ depends on $i$ or, equivalently, realizations of $W$ and $\bm{u}$, even after discarding the $O(N^{-\frac{1}{2}})$ term. Another related remark is that Eq.~(\ref{eq:15}) holds only in the limit of small $s^2$. Otherwise, the right-hand side may become negative even in the limit of large $N$, since $\left( V^n \bm{u} \right)_i$ follows a Gaussian distribution with a variance of $O(1)$ and thus can take an arbitrarily large value. 

In Fig.~\ref{fig:2}, the mean of $a_1(t)$ given $s(t-n)=0.1$ and the conditional probability density $P_{a_1(t)|s(t-9)}(a|c)$ are shown for single specific realizations of $W$ and $\bm{u}$. Here, we set $s^2=0.01$. The numerical results are obtained by first generating a single orbit of length $10^6$ time steps after discarding the initial $10^4$ time steps and then sampling the value of $a_1(t)$ with $0.1 \leq s(t-n) <0.11$ for each $n$. The theoretical values for $\mu_{n,0.1,1}$ and $\Sigma_{n,1}^2$ are calculated from Eqs.~(\ref{eq:11}), (\ref{eq:13}), and (\ref{eq:15}), where $W$ and $\bm{u}$ are the same as those used in the numerical simulation. We can see that the numerical results and the theoretical predictions agree well. 

\begin{figure*}[t]
\centering
\includegraphics[width=16cm]{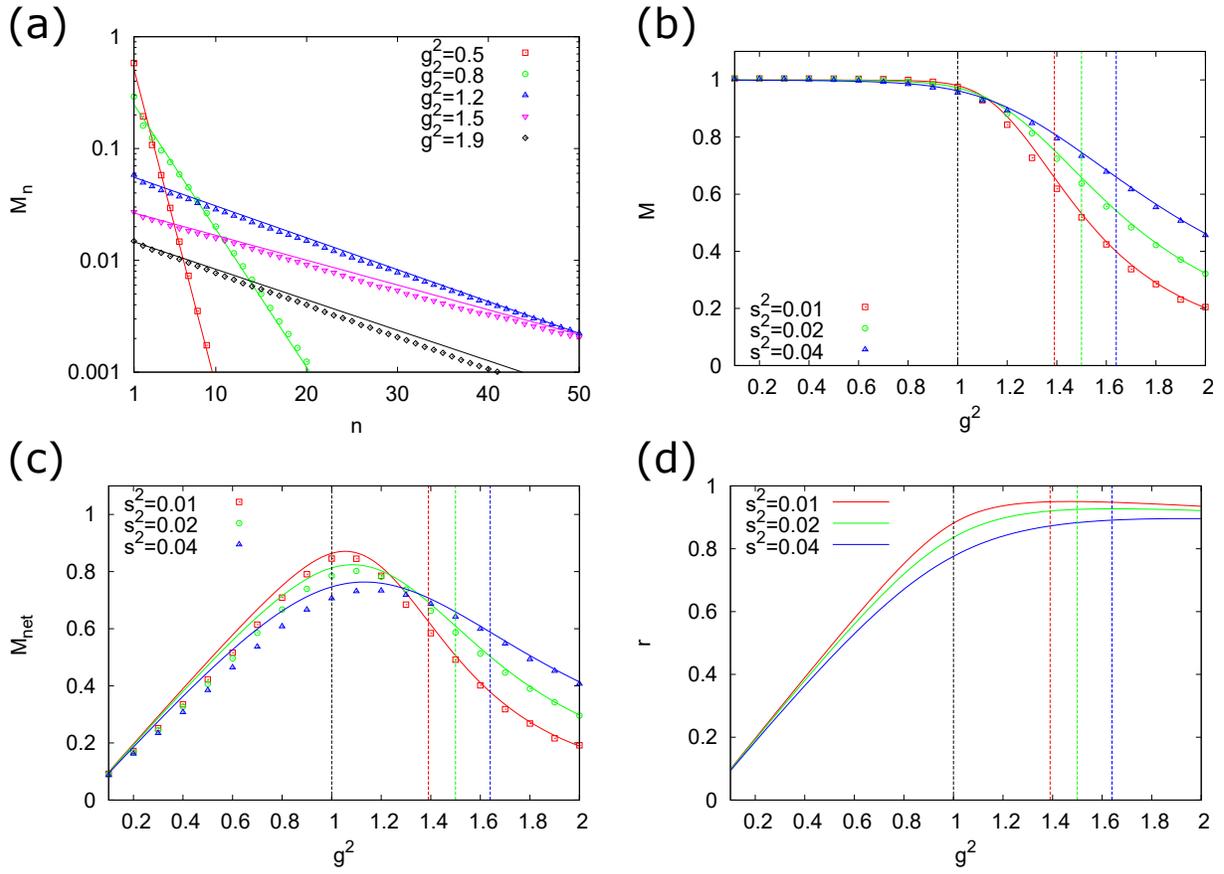}
\caption{
Memory capacity and network memory capacity of ESNs. Theoretical predictions and numerical results are compared for the population averages of (a) memory function $M_n$, (b) memory capacity $M$, and (c) network memory capacity $M_\mathrm{net}$. (d) Mean-field lines for the effective measurement of the nonlinear response $r$ (Eq.~(\ref{eq:19})). In (a), we set $s^2=0.01$. The vertical broken lines indicate the transition point between the ordered and chaotic regimes for the value of $s^2$ with the same color. 
}
\label{fig:3}
\end{figure*}

\begin{figure*}[t]
\centering
\includegraphics[width=16cm]{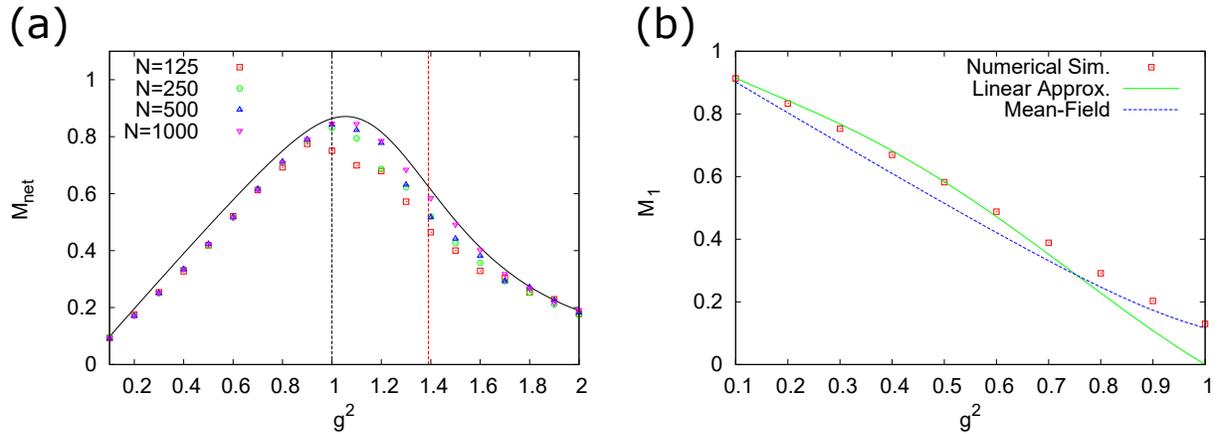}
\caption{
Linear approximation of $M_1$. (a) Numerical values of $\mathbf{E}\left[ M_\mathrm{net} \right]$ for different system sizes $N$. (b) Comparison among numerical results, linear approximation, and mean-field theory for $\mathbf{E}\left[ M_1 \right]$. We set $s^2=0.01$ in both panels. 
}
\label{fig:4}
\end{figure*}

Using Eqs.~(\ref{eq:13}) and (\ref{eq:15}), we obtain (Appendix~\ref{sec:c})
\begin{align}
\langle x_i(t)s(t-n) \rangle = \frac{\left( V^{n-1} \bm{u} \right)_i s^2}{\sqrt{1+\frac{\pi}{2}\Sigma^2}}
\label{eq:16}
\end{align}
for $n=1,2,\dots$. From Eqs.~(\ref{eq:7}) and (\ref{eq:16}), we have 
\begin{equation}
M_n = \frac{\left( V^{n-1} \bm{u} \right)_i^2 s^2}{\sigma^2 \left( 1+\frac{\pi}{2}\Sigma^2 \right)}
\label{eq:17}
\end{equation}
for $n=1,2,\dots$. The population average of $M_n$, which is equivalent to the average over realizations of $W$ and $\bm{u}$, is (Appendix \ref{sec:d})
\begin{equation}
\mathbf{E}\left[ M_n \right] = \frac{r^{n-1} s^2}{\sigma^2 \left( 1+\frac{\pi}{2}\Sigma^2 \right)} = \frac{r^n s^2}{g^2 \sigma^2}, 
\label{eq:18}
\end{equation}
where 
\begin{equation}
r=\frac{g^2}{1+\frac{\pi}{2}\Sigma^2}. 
\label{eq:19}
\end{equation}
The population average of $M$ is 
\begin{equation}
\mathbf{E}\left[ M \right]=\sum_{n=1}^\infty \mathbf{E}\left[ M_n \right] = \frac{r s^2}{g^2 \sigma^2 (1 - r)}. 
\label{eq:20}
\end{equation}
$M$ can be decomposed into two parts~\cite{Farkas2016,Schuecker2018}: the direct memory $M_1$ and the indirect memory through network $M_\textrm{net}:=M-M_1$. We call the latter the \textit{network memory capacity}. The population average of the latter is 
\begin{equation}
\mathbf{E}\left[ M_\mathrm{net} \right] = \frac{r^2 s^2}{g^2 \sigma^2 (1 - r)} = r \mathbf{E}\left[ M \right]. 
\label{eq:21}
\end{equation}

Figure~\ref{fig:3} (a) shows the population average of $M_n$ for different values of $g^2$ with $s^2=0.01$. The population averages of $M$ and $M_\mathrm{net}$ are shown in Fig.~\ref{fig:3} (b) and (c), respectively. $\mathbf{E}\left[ M_\mathrm{net} \right]$ peaks in the range $1 < g^2 < g_*^2$, where $g_*^2$ is the value of $g^2$ such that $\lambda=0$. The exact location of the maximum point depends on the value of $s^2$ and shifts to a larger value as $s^2$ increases. In the mean-field theory, $\mathbf{E}\left[ M_\mathrm{net} \right]$ is given as the product between $\mathbf{E}\left[ M \right]$ and $r$. Hence, its qualitative behavior can be understood from those of $\mathbf{E}\left[ M \right]$ and $r$ (Fig.~\ref{fig:3} (b) and (d), respectively). Since $\mathbf{E}\left[ M \right]$ is a measure of the linear short-term memory, it is expected to decrease as the nonlinearity of the system increases. On the other hand, $r$ can be interpreted as an effective measure of the nonlinear response of the system, which reaches saturation for sufficiently large $g^2$, since the activation function $f$ is a sigmoid function. Indeed, $r \to \frac{2}{\pi}$ as $g^2 \to \infty$, since $\sigma^2 \to 1$ as $g^2 \to \infty$ in Eq.~(\ref{eq:19}) (However, this cannot be seen from Fig.~\ref{fig:3} (d) because the range of $g^2$ shown is restricted upto $2$). 

The mean-field predictions and the numerical results agree well over the whole range of $g^2$ for $\mathbf{E}\left[ M \right]$. However, there is a clear discrepancy for $\mathbf{E}\left[ M_\mathrm{net} \right]$ in the ordered regime (Fig.~\ref{fig:3}(c)). This is due to the breakdown of the mean-field theory. That is, the assumption that $x_i(t)$ are independent and identically distributed and are also independent of $w_{ij}$ and $u_i$ is violated when the ESN dynamics is driven by input signals. Indeed, in a certain range of $g^2$ in the ordered regime ($0.2 < g^2 < 0.7$ in Fig.~\ref{fig:4} (a) where $s^2=0.01$), the numerically obtained values of $\mathbf{E}\left[ M_\mathrm{net} \right]$ do not approach the mean-field value as the system size $N$ increases. 
To understand the quantitative influence of the violation of the mean-field assumption on $\mathbf{E}\left[ M_\mathrm{net} \right]$, we consider the regime $g^2 \ll 1$, where the activation function $f$ can be approximated by the identity function $f(x)=x$. When $g^2 \ll 1$, we can approximately calculate $\mathbf{E}\left[ M_n \right]$ without the mean-field theory. Since both the mean-field theory and the linear approximation lead to $\mathbf{E}\left[ M \right]=1$ for $g^2 \ll 1$, the difference in $\mathbf{E}\left[ M_\mathrm{net} \right]$ is reduced to that in $\mathbf{E}\left[ M_1 \right]$. The linear approximation predicts (Appendix~\ref{sec:e}) 
\begin{equation}
\mathbf{E}\left[ M_1 \right] 
= s^2 \mathbf{E}\left[ \frac{1}{\sigma_i^2} \right] 
\simeq 1-g^2 + 2\frac{\left(1-g^2\right)^2}{1+g^2}g^4, 
\label{eq:22}
\end{equation}
where $\sigma_i^2$ is the stationary variance of $x_i(t)$. Note that we have $\mathbf{E}\left[ \sigma_i^2 \right]=\frac{s^2}{1-g^2}$ in both the mean-field theory and the linear approximation. Indeed, in the mean-field theory, Eq.~(\ref{eq:3}) reduces to $\sigma^2=\Sigma^2=g^2\sigma^2+s^2$ when $f(x)=x$. The equation for the linear approximation is derived in Appendix~\ref{sec:e} (Eq.~(\ref{eq:a14})). However, the former predicts $\mathbf{E}\left[ \frac{1}{\sigma_i^2} \right]=\frac{1}{\sigma^2}+O(N^{-\frac{1}{2}})$. When calculating $\mathbf{E}\left[ M_1 \right]$ for $g^2 \ll 1$, the mean-field theory fails to capture the variance of $\sigma_i^2$. 

We compare the values of $\mathbf{E}\left[ M_1 \right]$ obtained from the numerical simulation, the linear approximation, and the mean-field theory in Fig.~\ref{fig:4} (b). Although the mean-field line does not fit the numerical result, the linear approximation can explain it well.

\subsection{Mutual Information and Fisher Memory}
\label{sec:mi}

\begin{figure*}[t]
\centering
\includegraphics[width=16cm]{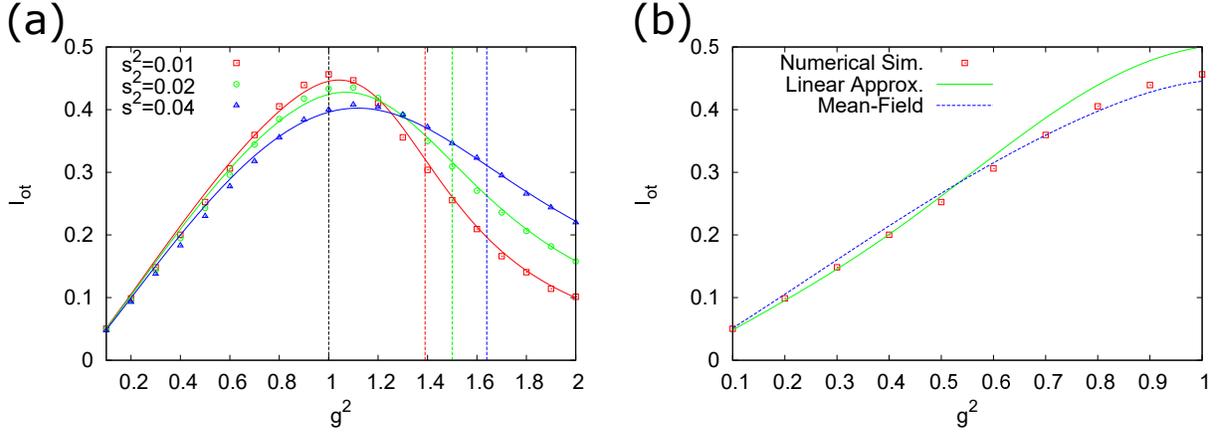}
\caption{
Network memory measure based on the mutual information between the future state and a one-time past input. (a) The theoretical predictions and the numerical results are compared for the population averages of $I_\mathrm{ot}$ (Eq.~(\ref{eq:25})). (b) Comparison among numerical results, linear approximation, and mean-field theory for $\mathbf{E}\left[ I_\mathrm{ot} \right]$ with $s^2=0.01$. 
}
\label{fig:5}
\end{figure*}

\begin{figure}[t]
\centering
\includegraphics[width=8cm]{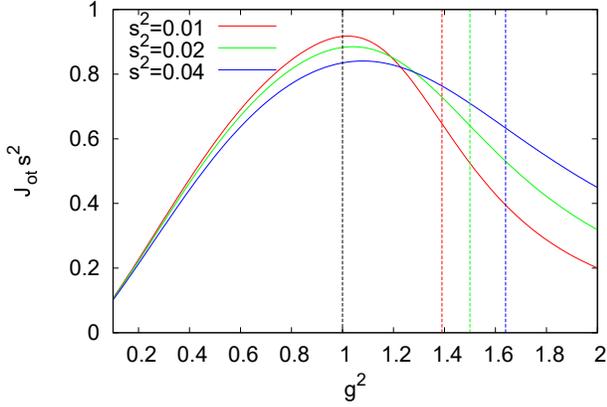}
\caption{
Population average of $J_\mathrm{ot}$ (Eq.~(\ref{eq:28})) calculated based on the mean-field theory under the same assumption as for Eq.~(\ref{eq:24}). To adjust the scale of the vertical axis, $J_\mathrm{ot}$ is multiplied by $s^2$. 
}
\label{fig:6}
\end{figure}

Once we obtain the conditional probability density $P_{a_i(t)|s(t-n)}(a|c)$ for $n=0,1,2,\dots$, we can immediately calculate the mutual information between $x_i(t+1)$ and $s(t-n)$ as 
\begin{equation}
I(x_i(t+1);s(t-n))=I(a_i(t);s(t-n)) \simeq \frac{1}{2} \log \frac{\Sigma^2}{\Sigma_{n,i}^2} 
\label{eq:23}
\end{equation}
when the mean-field assumption is valid. We would like to take the population average of Eq.~(\ref{eq:23}). Recall that we assumed $s^2 \ll 1$. Let us suppose $\Sigma^2=O(1)$ in the limit $s^2 \to 0$. In particular, this holds when $g^2>1$. Then, we can approximate the population average of $I(x_i(t+1);s(t-n))$ as 
\begin{align}
&\mathbf{E}\left[ I(x_i(t+1);s(t-n)) \right] \nonumber\\
&\simeq \frac{1}{2} \log \frac{\Sigma^2}{\mathbf{E}\left[ \Sigma_{n,i}^2 \right]} = \frac{1}{2} \log \frac{\Sigma^2}{\Sigma^2 - r^n s^2}. 
\label{eq:24}
\end{align}

Let us consider the summation of Eq.~(\ref{eq:23}) over $n=1,2,\dots$ defined by 
\begin{equation}
I_\mathrm{ot} = \sum_{n=1}^\infty I(x_i(t+1);s(t-n)), 
\label{eq:25}
\end{equation}
where the subscript $\mathrm{ot}$ indicates the mutual information between the future state and a one-time past input. Note that the $n=0$ term is not included in the summation. Thus, $I_\mathrm{ot}$ is a measure of network short-term memory analogous to $M_\mathrm{net}$ based on the mutual information. The population average $\mathbf{E}\left[ I_\mathrm{ot} \right]$ calculated based on the mean-field theory is shown in Fig.~\ref{fig:5}(a) and is compared with the numerical results. Since the direct numerical estimate of the mutual information between $x_i(t+1)$ and $s(t-n)$ for all $n=1,2,\dots,500$ is computationally hard, we estimated the mutual information from the correlation coefficient between $a_i(t)$ and $s(t-n)$ assuming that $P_{a_i(t)|s(t-n)}(a|c)$ is Gaussian, which is valid both in the linear regime and the mean-field regime. As in the case of $\mathbf{E}\left[ M_\mathrm{net} \right]$, $\mathbf{E}\left[ I_\mathrm{ot} \right]$ also takes a maximum value in the range $1 < g^2 < g_*^2$. 

As we have seen in the calculation of memory capacity, the mean-field theory is not applicable to the linear regime $g^2 \ll 1$. Indeed, although the discrepancy between the numerical results and the mean-field predictions appears to be small on the scale of Fig.~\ref{fig:5}(a), the calculation of $I_\mathrm{ot}$ based on the linear approximation (Appendix~\ref{sec:e}) provides much better fits to the numerical results than the mean-field theory for $g^2 \ll 1$, as shown in Fig.~\ref{fig:5}(b). 

\begin{figure*}[t]
\centering
\includegraphics[width=16cm]{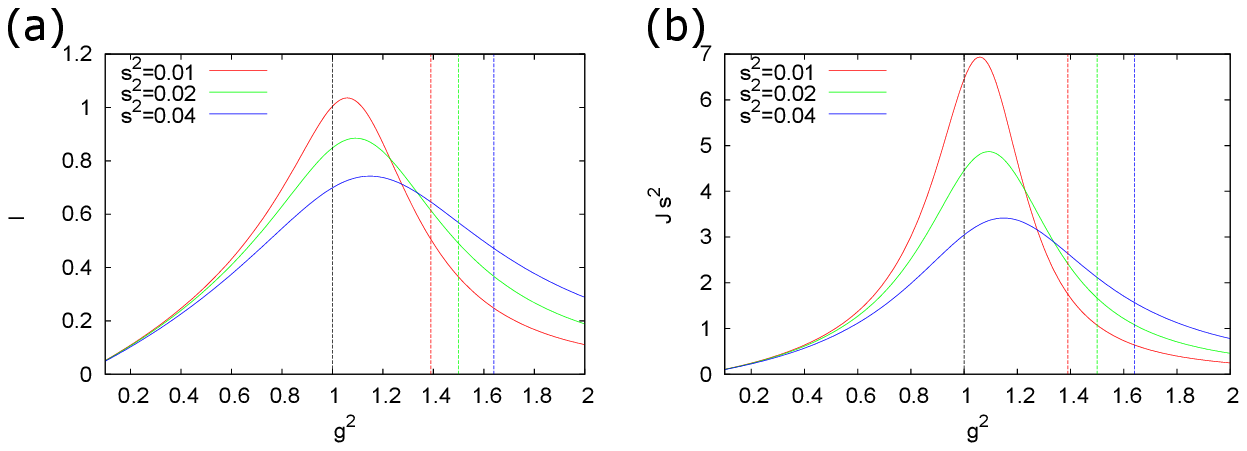}
\caption{
Population averages of (a) $I$ (Eq.~(\ref{eq:32})) and (b) $J \cdot s^2$ (Eq.~(\ref{eq:35}) multiplied by $s^2$) calculated from the mean-field theory under the same assumption as for Eq.~(\ref{eq:24}) are shown. 
}
\label{fig:7}
\end{figure*}

\begin{figure}[t]
\centering
\includegraphics[width=8cm]{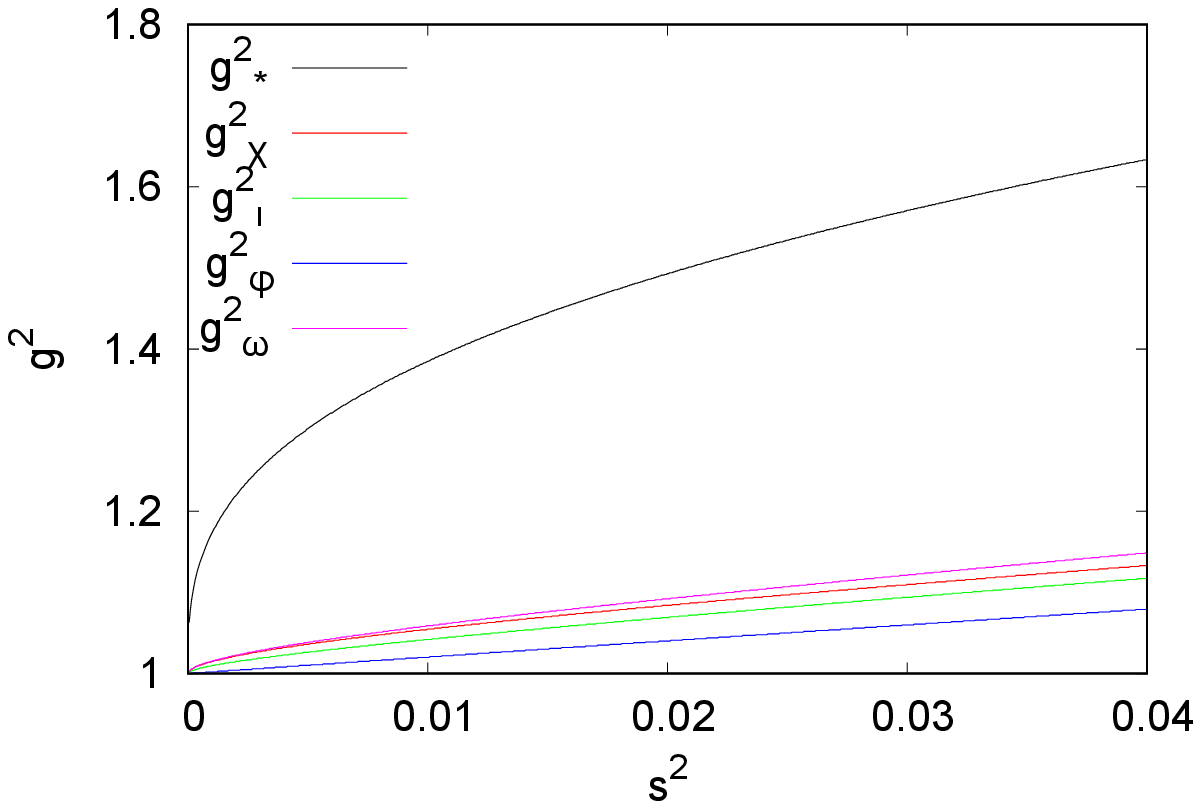}
\caption{The mean-field values of $g^2$ ($g_\chi^2$, $g_\iota^2$, $g_\phi^2$, and $g_\omega^2$) that attain the maxima of the five network memory measures ($\mathbf{E}\left[ M_\mathrm{net} \right]$, $\mathbf{E}\left[ I_\mathrm{ot} \right]$, $\mathbf{E}\left[ J_\mathrm{ot} \right]$, and $\mathbf{E}\left[ I \right]$ ($\mathbf{E}\left[ J \right]$), respectively) are shown as functions of $s^2$. Note that $\mathbf{E}\left[ I \right]$ and $\mathbf{E}\left[ J \right]$ peak at the same value of $g^2$. The critical line $g^2=g_*^2$ is also shown. 
}
\label{fig:8}
\end{figure}

\begin{figure}[t]
\centering
\includegraphics[width=8cm]{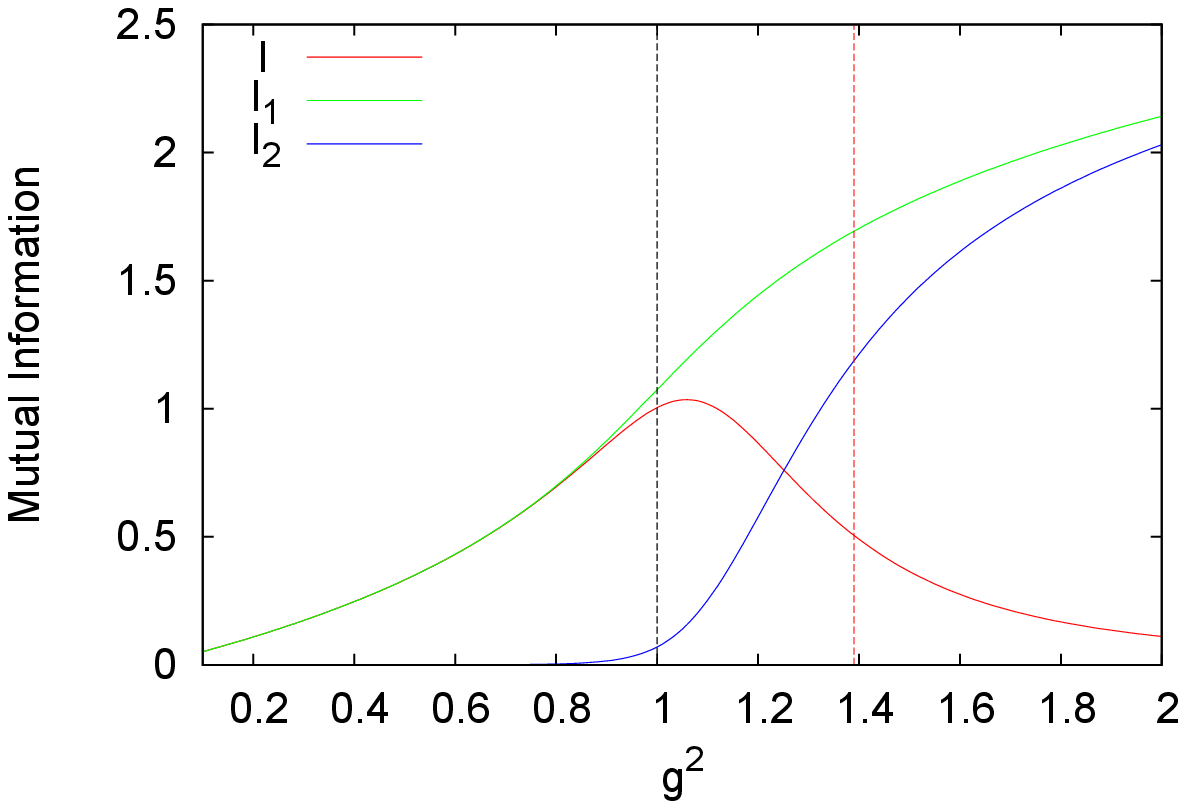}
\caption{$I=\lim_{n \to \infty}I(x_i(t+1);\bm{s}_{1:n}(t))$ can be represented as the difference between $I_1=I(x_i(t+1);\bm{x}(t))$ and $I_2=\lim_{n \to \infty}I(x_i(t+1);\bm{x}(t) | \bm{s}_{1:n}(t))$ (Eq.~(\ref{eq:36})). The population-averaged values of these quantities calculated from the mean-field theory under the same assumption as for Eq.~(\ref{eq:24}) are shown for $s^2=0.01$. 
}
\label{fig:9}
\end{figure}

Another familiar information-theoretic memory measure is the Fisher information~\cite{Cover2006, Ganguli2008}. Here, we regard the past input $s(t-n)$ as a parameter and consider the Fisher information with respect to the conditional probability density $P_{x_i(t+1)|s(t-n)}(x|c)$, namely, information about $s(t-n)$ contained in $x_i(t+1)$. Since the activation function $f$ is invertible and the Fisher information is invariant under an invertible transformation of stochastic variables, we can use $P_{a_i(t)|s(t-n)}(a|c)$ to calculate the Fisher information. When $s^2 \ll 1$ and the mean-field assumption is valid, the Fisher information for $s(t-n)$ contained in $x_i(t+1)$ is calculated as 
\begin{align}
&J(n) \nonumber\\
&= \int_{-\infty}^\infty da \, \left( \frac{\partial}{\partial c} \log P_{a_i(t)|s(t-n)}(a|c) \right)^2 P_{a_i(t)|s(t-n)}(a|c) \nonumber\\
&\simeq \frac{\left( \frac{\partial \mu_{n,c,i}}{\partial c} \right)^2}{\Sigma_{n,i}^2} = \frac{\left( V^n \bm{u} \right)_i^2}{\Sigma_{n,i}^2}. 
\label{eq:26}
\end{align}
The population average of Eq.~(\ref{eq:26}) can be approximately obtained under the same assumption as for Eq.~(\ref{eq:24}) as follows:
\begin{equation}
\mathbf{E}\left[ J(n) \right] \simeq \frac{\mathbf{E}\left[ \left( V^n \bm{u} \right)_i^2 \right] }{\mathbf{E}\left[ \Sigma_{n,i}^2 \right]} = \frac{r^n}{\Sigma^2-r^n s^2}. 
\label{eq:27}
\end{equation}
We define the network Fisher memory with respect to a one-time past input by 
\begin{equation}
J_\mathrm{ot} = \sum_{n=1}^\infty J(n). 
\label{eq:28}
\end{equation}
As in the case of $I_\mathrm{ot}$, we exclude the $n=0$ term representing direct memory from the sum in the right-hand side of Eq.~(\ref{eq:28}). The population average $\mathbf{E}\left[ J_\mathrm{ot} \right]$ calculated based on the mean-field theory under the same assumption as for Eq.~(\ref{eq:24}) is shown in Fig.~\ref{fig:6}. $\mathbf{E}\left[ J_\mathrm{ot} \right]$ behaves qualitatively similarly to $\mathbf{E}\left[ M_\mathrm{net} \right]$ and $\mathbf{E}\left[ I_\mathrm{ot} \right]$ at least in the range where the mean-field theory is valid. Note that there is a close relationship between the mean-field predictions of memory function, mutual information, and Fisher information through Eqs.~(\ref{eq:18}), (\ref{eq:24}) and (\ref{eq:27}): $\mathbf{E}\left[ I(x_i(t+1);s(t-n)) \right] = \frac{1}{2}\log \left( 1 - \left( 1-\frac{s^2}{\Sigma^2} \right) \mathbf{E}\left[ M_n \right] \right) = \frac{1}{2} \log \left( 1+\mathbf{E}\left[ J(n) \right]s^2 \right)$. 

The derivation of the conditional probability density $P_{a_i(t)|s(t-n)}(a|c)$ under the mean-field assumption can be extended in a straightforward manner to the conditioning on a set of past inputs at multiple time steps. In particular, the conditional probability density $P_{a_i(t)|\bm{s}_{1:n}(t)}(a|\bm{c})$ can be approximated as a Gaussian distribution with mean $\mu_{1:n,\bm{c},i}$ and variance $\Sigma_{1:n,i}^2$ where $\bm{s}_{1:n}(t)=(s(t-1),s(t-2),\dots,s(t-n))$, $\bm{c}=(c_1,c_2,\dots,c_n)$, 
\begin{equation}
\mu_{1:n,\bm{c},i} = \sum_{k=1}^n \left(V^k \bm{u} \right)_i c_k
\label{eq:29}
\end{equation}
and 
\begin{equation}
\Sigma_{1:n,i}^2 = \Sigma^2 - \sum_{k=1}^n \left(V^k \bm{u} \right)_i^2 s^2. 
\label{eq:30}
\end{equation}

An alternative network memory measure to $I_\mathrm{ot}$ is the limit $n \to \infty$ of the mutual information between $x_i(t+1)$ and $\bm{s}_{1:n}(t)$. When the mean-field assumption is valid, it is given by 
\begin{align}
I =& \lim_{n \to \infty}I(x_i(t+1);\bm{s}_{1:n}(t)) \nonumber\\
=& \lim_{n \to \infty}I(a_i(t);\bm{s}_{1:n}(t)) \simeq \frac{1}{2} \log \frac{\Sigma^2}{\Sigma_{1:\infty,i}^2}, 
\label{eq:31}
\end{align}
where $\Sigma_{1:\infty,i}^2=\Sigma^2 - \sum_{k=1}^\infty \left(V^k \bm{u} \right)_i^2 s^2$. Under the same assumption as for Eq.~(\ref{eq:24}), its population average is approximately given by 
\begin{equation}
\mathbf{E}\left[ I \right] \simeq \frac{1}{2} \log \frac{(1-r)\Sigma^2}{(1-r)\Sigma^2 - r s^2}. 
\label{eq:32}
\end{equation}

Similarly, we can consider an alternative network memory measure based on the Fisher information matrix with respect to $P_{a_i(t)|\bm{s}_{1:n}(t)}(a|\bm{c})$. When the mean-field assumption is valid, the $(k,l)$-th element of the Fisher information matrix is given by 
\begin{equation}
J_{kl} \simeq \frac{\frac{\partial \mu_{1:n,\bm{c},i}}{\partial c_k}\frac{\partial \mu_{1:n,\bm{c},i}}{\partial c_l}}{\Sigma_{1:n,i}^2} = \frac{\left(V^k \bm{u} \right)_i \left(V^l \bm{u} \right)_i}{\Sigma_{1:n,i}^2}. 
\label{eq:33}
\end{equation}
The Fisher memory with respect to the whole past input history is defined by 
\begin{equation}
J = \lim_{n \to \infty}\sum_{k=1}^n J_{kk}
\label{eq:34}
\end{equation}
and its approximate population average is found to be 
\begin{equation}
\mathbf{E}\left[ J \right] \simeq \frac{r}{(1-r)\Sigma^2 - r s^2} 
\label{eq:35}
\end{equation}
under the same assumption as for Eq.~(\ref{eq:24}). Note that $\mathbf{E}\left[ I \right]$ and $\mathbf{E}\left[ J \right]$ are related by $\mathbf{E}\left[ I \right] \simeq \frac{1}{2}\log \left( 1 + \mathbf{E}\left[ J \right] s^2 \right)$. Figure~\ref{fig:7} shows Eqs.~(\ref{eq:32}) and (\ref{eq:35}) for $s^2=0.01, 0.02$, and $0.04$. Both $\mathbf{E}\left[ I \right]$ and $\mathbf{E}\left[ J \right]$ take maximum values at points close to those for $\mathbf{E}\left[ M_\mathrm{net} \right]$, $\mathbf{E}\left[ I_\mathrm{ot} \right]$, and $\mathbf{E}\left[ J_\mathrm{ot} \right]$ as long as the mean-field theory is valid. Figure~\ref{fig:8} summarizes the maximum points of these network memory measures in the range $0<s^2\leq 0.04$ obtained from the mean-field theory. 

Finally, we remark on the behavior of $I$. Because $x_i(t+1)$ and $\bm{x}(t)$ are conditionally independent given $\bm{s}_{1:n}(t)$, we obtain 
\begin{align}
&I(x_i(t+1);\bm{x}(t)) \nonumber\\
&= I(x_i(t+1);\bm{s}_{1:n}(t)) + I(x_i(t+1);\bm{x}(t) | \bm{s}_{1:n}(t)). 
\label{eq:36}
\end{align}
The first term on the right-hand side of Eq.~(\ref{eq:36}) becomes $I$ by taking the limit $n \to \infty$. The second term $I(x_i(t+1);\bm{x}(t) | \bm{s}_{1:n}(t))$ will be negligible when the system is driven by input signals ($g^2 \ll 1$). On the other hand, the chaotic dynamics dominates as $g^2 \to \infty$ and $I(x_i(t+1);\bm{x}(t) | \bm{s}_{1:n}(t))$ will approach $I(x_i(t+1);\bm{x}(t))$. Thus, as $g^2$ varies from $0$ to $\infty$, $I$ will increase together with $I(x_i(t+1);\bm{x}(t))$ in the ordered regime but will decrease in the sufficiently chaotic regime. Hence, $I$ is expected to take a maximum value between the two extremes. In Fig.~\ref{fig:9}, the population-averaged values of the three terms in Eq.~(\ref{eq:36}) in the limit $n \to \infty$ calculated from the mean-field theory under the same assumption as for Eq.~(\ref{eq:24}) are shown, where $I(x_i(t+1);\bm{x}(t))=\frac{1}{2} \log \frac{\Sigma^2}{s^2}$ due to the mean-field assumption. 

\section{Discussion}
\label{sec:discussion}
The three network memory measures studied in this paper take maximum values in the ordered regime for ESNs with small input signals. The value of $g^2$ that attains the maximum is always greater than $1$, which is the boundary between the ordered and chaotic regimes in the corresponding autonomous system. However, it is far from the critical $g_*^2$ (Fig.~\ref{fig:8}). In previous work, it was argued that the maximal Fisher information can be used to detect the edge of chaos~\cite{Prokopenko2011,Livi2017}. Our results suggest that such an approach is not necessarily effective for driven dynamical systems. 

In the context of physical reservoir computing~\cite{Larger2012,Larger2017,Torrejon2017,Tsunegi2019,Nakajima2018,Fujii2017}, it is generally difficult to tune the parameters of a given physical system for optimal computational performance. An alternative method is to choose an optimal input strength. Although the analysis presented in this paper assumes that $s^2$ is small, our results theoretically suggest that tuning the input strength is meaningful. For example, the value of $\mathbf{E}\left[ M_\mathrm{net} \right]$ for $s^2=0.02$ is greater than those for $s^2=0.01$ and $s^2=0.04$ around $g^2 \approx 1.3$ in Fig.~\ref{fig:3} (c). 

Toyoizumi and Abbott~\cite{Toyoizumi2011} analytically showed that the signal-to-noise ratio of ESNs decreases rapidly on the left side of the criticality $g^2=1$ when inputs are absent, but decreases much slowly on the right side. They suggested that high computational performance can be achieved without fine tuning in the latter. Our results confirm this expectation because all of the short-term memory measures in the presence of inputs peak near $g^2=1$ in the region $1 < g^2 < g_*^2$. 

In general, dynamical regimes of autonomous systems beyond stable fixed points are candidates for computational resources.  For example, RNNs with sinusoidal activation functions achieve a high computational performance in the non-chaotic window regions after transition to chaos occurs in their autonomous dynamics~\cite{Marquez2018}. An online supervised learning algorithm for RNNs proposed by Sussillo and Abbott~\cite{Sussillo2009} exhibits its best performance when their autonomous dynamics is adjusted to the chaotic region not far from the critical point where chaotic dynamics can be suppressed by input signals. Schuecker et al.~\cite{Schuecker2018} showed that the network memory capacity for continuous-time nonlinear RNNs peaks in the ordered regime with $g^2>1$, which is consistent to our result. They argued that the dynamic suppression of chaos (DSC), which results from the fact that the onset of local instability precedes that of asymptotic instability, contributes to optimal information processing. However, DSC cannot occur in discrete-time ESNs where the two onsets coincide. In ESNs, the shift of the critical $g_*^2$ toward the chaotic regime induced by input signals is solely due to a mechanism called the static suppression of chaos (SSC), which increases the frequency with which an orbit visits the contracting region of the phase space. Unlike SSC, DSC is conjectured to occur based on fast switching among different unstable directions caused by input signals~\cite{Schuecker2018}. However, ESNs with leaky neurons~\cite{Jaeger2007} are expected to exhibit DSC~\cite{Schuecker2018}. Future analyses of network memory measures for leaky ESNs based on the presented theory could deepen the understanding of the relationship between DSC and the information processing ability of dynamical systems. 

It has been suggested that there exists a trade-off between nonlinearity of dynamical systems and their memory capacity~\cite{Dambre2012}. Inubushi and Yoshimura~\cite{Inubushi2017} theoretically investigated the trade-off in terms of how nonlinearity degrades small initial differences of input signals. The mean-field theory presented in this paper makes it possible to study the trade-off when input strength is small by directly calculating the nonlinear memory capacity proposed by Damble et al.~\cite{Dambre2012}. Performing the detailed calculation is also left as future work. 

\appendix

\section{Derivation of Eq.~(\ref{eq:10})}
\label{sec:a}
We have 
\begin{align}
&\left\langle \frac{\delta a_i(t)}{\delta s(t-n)} \right\rangle \nonumber\\
&= \sum_{j_1,\dots,j_n} w_{i j_1} \dots w_{j_{n-1}j_n}u_{j_n}\left\langle \prod_{k=1}^n f'(a_{j_k}(t-k)) \right\rangle. 
\label{eq:a0}
\end{align}
Since $a_{j_k}(t-k)$ $(k=1,2,\dots,n)$ are independent and $a_{j_k}(t-k) \sim N(0,\Sigma^2)$ by the mean-field assumption, we have
\begin{align}
\left\langle \prod_{k=1}^n f'(a_{j_k}(t-k)) \right\rangle 
&= \prod_{k=1}^n  \left\langle f'(a_{j_k}(t-k)) \right\rangle \nonumber\\
&= \left( \int_{-\infty}^\infty da \, f'(a) \frac{\exp\left( - \frac{a^2}{2\Sigma^2} \right)}{\sqrt{2\pi \Sigma^2}} \right)^n \nonumber\\
&= \left( \frac{1}{\sqrt{1+\frac{\pi}{2}\Sigma^2}} \right)^n, 
\label{eq:a1}
\end{align}
where $f'(a)=e^{-\frac{\pi}{4}a^2}$ is used for $f(a)=\textrm{erf}(\frac{\sqrt{\pi}}{2}a)$. Equation~(\ref{eq:10}) follows from Eqs.~(\ref{eq:11}), (\ref{eq:a0}) and (\ref{eq:a1}).

\section{Derivation of Eq.~(\ref{eq:12})}
\label{sec:b}
We have 
\begin{align}
&\left\langle \frac{\delta a_i(t)}{\delta s(t-n)} s(t-n) \right\rangle \nonumber\\
&= \sum_{j_1,\dots,j_n} w_{i j_1} \dots w_{j_{n-1}j_n}u_{j_n} \left\langle \left( \prod_{k=1}^n f'(a_{j_k}(t-k)) \right) s(t-n) \right\rangle. 
\label{eq:a2}
\end{align}
It is sufficient to show 
\begin{equation}
\left\langle \left( \prod_{k=1}^n f'(a_{j_k}(t-k)) \right) s(t-n) \right\rangle = 0.
\label{eq:a3}
\end{equation}
We set 
\begin{equation}
g_n(y,z_n)=y+u_{j_n}z_n
\label{eq:a4}
\end{equation}
and 
\begin{widetext}
\begin{align}
g_{n-k}(y,z_{n-k},\dots,z_n) 
= \sum_{j_{n-k+1}} w_{j_{n-k}j_{n-k+1}} f(g_{n-k+1}(y,z_{n-k+1},\dots,z_n)) + u_{j_{n-k}}z_{n-k}
\label{eq:a5}
\end{align}
\end{widetext}
for $k=1,\dots,n-1$. Let us introduce 
\begin{equation}
G(z_1,\dots,z_n)=z_n \int_{-\infty}^\infty Dy \, \prod_{k=1}^n f'(g_k(y,z_k,\dots,z_n)),
\label{eq:a6}
\end{equation}
where $Dy=\frac{dy}{\sqrt{2\pi g^2 \sigma^2}}e^{-\frac{y^2}{2 g^2 \sigma^2}}$. The left-hand side of Eq.~(\ref{eq:a3}) can be written as 
\begin{align}
&\left\langle \left( \prod_{k=1}^n f'(a_{j_k}(t-k)) \right) s(t-n) \right\rangle \nonumber\\
&= \int_{-\infty}^\infty Dz_1 \dots \int_{-\infty}^\infty Dz_n \, G(z_1,\dots,z_n),
\label{eq:a7}
\end{align}
where $Dz_k=\frac{dz_k}{\sqrt{2\pi s^2}}e^{-\frac{z_k^2}{2 s^2}}$. We shall show that $G$ is an odd function with respect to $(z_1,\dots,z_n)$, namely, 
\begin{equation}
G(-z_1,\dots,-z_n)=-G(z_1,\dots,z_n) 
\label{eq:a8}
\end{equation}
holds. This yields the desired result. First, note that $g_{n-k}(y,z_{n-k},\dots,z_n)$ is odd with respect to $(y,z_{n-k},\dots,z_n)$ for $k=0,1,\dots,n-1$. Namely, we have 
\begin{equation}
g_{n-k}(-y,-z_{n-k},\dots,-z_n) = -g_{n-k}(y,z_{n-k},\dots,z_n). 
\label{eq:a9}
\end{equation}
Indeed, Eq.~(\ref{eq:a9}) can be proved by mathematical induction. First, $g_n(-y,-z_n)=-y-u_{j_n}z_n=-g_n(y,z_n)$ for $k=0$. Assume that $g_{n-k}(-y,-z_{n-k},\dots,-z_n)=-g_{n-k}(y,z_{n-k},\dots,z_n)$ for $k=0,1,\dots,n-2$. Then, we obtain 
\begin{widetext}
\begin{align}
g_{n-(k+1)}(-y,-z_{n-(k+1)},\dots,-z_n) &= \sum_{j_{n-k}} w_{j_{n-(k+1)}j_{n-k}} f(g_{n-k}(-y,-z_{n-k},\dots,-z_n)) - u_{j_{n-(k+1)}}z_{n-(k+1)} \nonumber\\
&= \sum_{j_{n-k}} w_{j_{n-(k+1)}j_{n-k}} f(-g_{n-k}(y,z_{n-k},\dots,z_n)) - u_{j_{n-(k+1)}}z_{n-(k+1)} \nonumber\\
&= - \sum_{j_{n-k}} w_{j_{n-(k+1)}j_{n-k}} f(g_{n-k}(y,z_{n-k},\dots,z_n)) - u_{j_{n-(k+1)}}z_{n-(k+1)} \nonumber\\
&= -g_{n-(k+1)}(y,z_{n-(k+1)},\dots,z_n), 
\label{eq:a10}
\end{align}
\end{widetext}
where we applied the induction hypothesis for the third equality and we used the fact that $f$ is an odd function for the fourth equality. Now, Eq.~(\ref{eq:a8}) is obtained by 
\begin{align}
&G(-z_1,\dots,-z_n) \nonumber\\
&= -z_n \int_{-\infty}^\infty Dy \, \prod_{k=1}^n f'(g_k(y,-z_k,\dots,-z_n)) \nonumber\\
&= -z_n \int_{-\infty}^\infty Dy \, \prod_{k=1}^n f'(g_k(-y,-z_k,\dots,-z_n)) \nonumber\\
&= -z_n \int_{-\infty}^\infty Dy \, \prod_{k=1}^n f'(-g_k(y,z_k,\dots,z_n)) \nonumber\\
&= -z_n \int_{-\infty}^\infty Dy \, \prod_{k=1}^n f'(g_k(y,z_k,\dots,z_n)) \nonumber\\
&= -G(z_1,\dots,z_n),
\label{eq:a11}
\end{align}
where we used Eq.~(\ref{eq:a9}) for the third equality and the fact that $f'$ is an even function for the fourth equality.

\section{Derivation of Eq.~(\ref{eq:16})}
\label{sec:c}
We have 
\begin{align}
&\langle x_i(t)s(t-n) \rangle \nonumber\\
&= \int_{-\infty}^{\infty} da \, \int_{-\infty}^{\infty} dc \, P_{a_i(t-1) | s(t-n)}(a|c) P_{s(t-n)}(c) f(a)c \nonumber\\
&= \frac{\left( V^{n-1} \bm{u} \right)_i s^2}{\sqrt{2\pi}\Sigma^3} \int_{-\infty}^\infty da \, e^{-\frac{a^2}{2\Sigma^2}} f(a) a. 
\label{eq:a20}
\end{align}
By the change of variable $y=\frac{a}{\Sigma}$, the integral on the right-hand side of Eq.~(\ref{eq:a20}) can be calculated as 
\begin{align}
\int_{-\infty}^\infty da \, e^{-\frac{a^2}{2\Sigma^2}} f(a) a 
&= \Sigma^2 \int_{-\infty}^\infty dy \, e^{-\frac{y^2}{2}} f(\Sigma y) y \nonumber\\
&= \Sigma^2 \int_{-\infty}^\infty dy \, \frac{d}{dy}\left( - e^{-\frac{y^2}{2}}\right)f(\Sigma y) \nonumber\\
&= \Sigma^3 \int_{-\infty}^\infty dy \, e^{-\frac{\left( 1+\frac{\pi}{2}\Sigma^2 \right)y^2}{2}} \nonumber\\
&= \frac{\sqrt{2\pi}\Sigma^3}{\sqrt{1+\frac{\pi}{2}\Sigma^2}}. 
\label{eq:a21}
\end{align}
From Eqs.~(\ref{eq:a20}) and (\ref{eq:a21}), we obtain Eq.~(\ref{eq:16}). 

\section{Derivation of Eq.~(\ref{eq:18})}
\label{sec:d}
Since $\left( V^n \bm{u} \right)_i^2 = \left( \frac{1}{1+\frac{\pi}{2}\Sigma^2} \right)^n \left( W^n \bm{u} \right)_i^2$, it is sufficient to show $\mathbf{E} \left[ \left( W^n \bm{u} \right)_i^2 \right] = g^{2n} + O(\frac{1}{N})$ for $n=0,1,\dots$. 

Let $w_{ij}^{(n)}$ be the $(i,j)$-th element of $W^n$. We have 
\begin{align}
\mathbf{E} \left[ \left( W^n \bm{u} \right)_i^2 \right]
&= \sum_{j,k} \mathbf{E} \left[ w_{ij}^{(n)} w_{ik}^{(n)} \right] \mathbf{E} \left[ u_j u_k \right] \nonumber\\
&= \sum_j \mathbf{E} \left[ \left( w_{ij}^{(n)} \right)^2 \right], 
\label{eq:a22}
\end{align}
where we used $\mathbf{E} \left[ u_j u_k \right]=\delta_{jk}$ and $\delta_{jk}$ is the Kronecker delta. The population average of $\left( w_{ij}^{(n)} \right)^2$ is given by 
\begin{equation}
\mathbf{E} \left[ \left( w_{ij}^{(n)} \right)^2 \right] 
= \frac{g^{2n}}{N} + O\left(\frac{1}{N^2}\right), 
\label{eq:a23}
\end{equation}
because $w_{ij} \sim N(0,\frac{g^2}{N})$ are independent. Thus, we obtain 
\begin{equation}
\mathbf{E} \left[ \left( W^n \bm{u} \right)_i^2 \right]=g^{2n} + O\left(\frac{1}{N}\right). 
\label{eq:a24}
\end{equation}

\section{Linear Approximation}
\label{sec:e}
When $g^2 \ll 1$, we can approximate $f(a)=a$ and obtain 
\begin{equation}
\bm{x}(t+1) = \sum_{n=0}^\infty s(t-n)W^n \bm{u}. 
\label{eq:a12}
\end{equation}
Thus, 
\begin{equation}
\sigma_i^2 = \langle x_i(t+1)^2 \rangle = s^2 \sum_{n=0}^\infty \left( W^n \bm{u} \right)_i^2
\label{eq:a13}
\end{equation}
holds. Ignoring the $O(\frac{1}{N})$ terms, the mean and variance of $\left( W^n \bm{u} \right)_i^2$ for $n \geq 1$ are given by $g^{2n}$ and $2g^{4n}$, respectively, because $w_{jk}$ and $u_j$ are independent and $w_{jk} \sim N(0,\frac{g^2}{N})$ and $u_j=\pm 1$. The population average of $\sigma_i^2$ is given by 
\begin{equation}
\mathbf{E} \left[ \sigma_i^2 \right] = s^2 \sum_{n=0}^\infty g^{2n} = \frac{s^2}{1-g^2}. 
\label{eq:a14}
\end{equation}
Its variance is 
\begin{align}
\mathrm{Var} \left[ \sigma_i^2 \right] 
&= s^4 \left( \mathbf{E} \left[ \left( \sum_{n=0}^\infty \left( W^n \bm{u} \right)_i^2 \right)^2 \right] - \mathbf{E} \left[ \sum_{n=0}^\infty \left( W^n \bm{u} \right)_i^2 \right]^2 \right) \nonumber\\
&= s^4 \sum_{n=0}^\infty \mathrm{Var} \left[ \left( W^n \bm{u} \right)_i^2 \right] + O\left(\frac{1}{N}\right) \nonumber\\
&= 2s^4 \sum_{n=1}^\infty g^{4n} = \frac{2 g^4 s^4}{1-g^4}, 
\label{eq:a15}
\end{align}
where we used $\mathbf{E}\left[ \left( W^m \bm{u} \right)_i^2 \left( W^n \bm{u} \right)_i^2 \right] = \mathbf{E}\left[ \left( W^m \bm{u} \right)_i^2 \right] \mathbf{E}\left[ \left( W^n \bm{u} \right)_i^2 \right] + O(\frac{1}{N})$ for $m \neq n$. From Eqs.~(\ref{eq:a14}) and (\ref{eq:a15}), we obtain 
\begin{align}
\mathbf{E} \left[ \frac{1}{\sigma_i^2} \right] 
&\simeq \frac{1}{\mathbf{E} \left[ \sigma_i^2 \right]} \left( 1 + \frac{\mathrm{Var} \left[ \sigma_i^2 \right]}{\mathbf{E} \left[ \sigma_i^2 \right]^2} \right) \nonumber\\
&= \frac{1-g^2}{s^2} \left( 1 + 2 \frac{1-g^2}{1+g^2}g^4 \right). 
\label{eq:a16}
\end{align}
Equation~(\ref{eq:22}) follows from Eq.~(\ref{eq:a16}). 

Similarly, we can compute the mutual information between $x_i(t+1)$ and $s(t-n)$ in the linear regime $g^2 \ll 1$ as follows. Let $X:= \sum_{m=0}^\infty \left( W^m \bm{u} \right)_i^2$ and $X_n:=\sum_{m=0,m \neq n}^\infty \left( W^m \bm{u} \right)_i^2$. We have 
\begin{equation}
\mathbf{E} \left[ I(x_i(t+1);s(t-n))\right] = \frac{1}{2} \mathbf{E} \left[ \log \frac{X}{X_n} \right]. 
\label{eq:a17}
\end{equation}
Since $\mathbf{E} \left[ \log X \right]$ can be approximated as 
\begin{equation}
\mathbf{E} \left[ \log X \right] \simeq \log \mathbf{E} \left[ X \right] - \frac{1}{2}\frac{\mathrm{Var}\left[ X \right]}{\mathbf{E} \left[ X \right]^2} 
\label{eq:a18}
\end{equation}
and a similar approximation can be obtained for $\mathbf{E} \left[ \log X_n \right]$, we obtain 
\begin{align}
&\mathbf{E} \left[ I(x_i(t+1);s(t-n))\right] \nonumber\\
&\simeq \frac{1}{2} \log\frac{\mathbf{E} \left[ X \right]}{\mathbf{E} \left[ X_n \right]} + \frac{1}{4} \left( \frac{\mathrm{Var}\left[ X_n \right]}{\mathbf{E} \left[ X_n \right]^2} - \frac{\mathrm{Var}\left[ X \right]}{\mathbf{E} \left[ X \right]^2} \right), 
\label{eq:a19}
\end{align}
where $\mathbf{E} \left[ X \right] = \frac{1}{1-g^2}$, $\mathbf{E} \left[ X_n \right] = \frac{1}{1-g^2} - g^{2n}$, $\mathrm{Var}\left[ X \right] = \frac{2g^4}{1-g^4}$ and $\mathrm{Var}\left[ X_n \right] = \frac{2g^4}{1-g^4} - 2g^{4n}$. $\mathbf{E}\left[ I_\mathrm{ot} \right]$ can be computed by summing Eq.~(\ref{eq:a19}) over $n \geq 1$.

\begin{acknowledgments}
The authors are grateful to the anonymous reviewers for their comments and suggestions that improved the manuscript. TH was supported by JSPS KAKENHI Grant Number JP18K03423. KN was supported by JSPS KAKENHI Grant Number JP18H05472 and by MEXT Quantum Leap Flagship Program (MEXT Q-LEAP) Grant Number JPMXS0118067394. This work is partially based on results obtained from a project commissioned by the New Energy and Industrial Technology Development Organization (NEDO).
\end{acknowledgments}


\begin{thebibliography}{35}%
\makeatletter
\providecommand \@ifxundefined [1]{%
 \@ifx{#1\undefined}
}%
\providecommand \@ifnum [1]{%
 \ifnum #1\expandafter \@firstoftwo
 \else \expandafter \@secondoftwo
 \fi
}%
\providecommand \@ifx [1]{%
 \ifx #1\expandafter \@firstoftwo
 \else \expandafter \@secondoftwo
 \fi
}%
\providecommand \natexlab [1]{#1}%
\providecommand \enquote  [1]{``#1''}%
\providecommand \bibnamefont  [1]{#1}%
\providecommand \bibfnamefont [1]{#1}%
\providecommand \citenamefont [1]{#1}%
\providecommand \href@noop [0]{\@secondoftwo}%
\providecommand \href [0]{\begingroup \@sanitize@url \@href}%
\providecommand \@href[1]{\@@startlink{#1}\@@href}%
\providecommand \@@href[1]{\endgroup#1\@@endlink}%
\providecommand \@sanitize@url [0]{\catcode `\\12\catcode `\$12\catcode
  `\&12\catcode `\#12\catcode `\^12\catcode `\_12\catcode `\%12\relax}%
\providecommand \@@startlink[1]{}%
\providecommand \@@endlink[0]{}%
\providecommand \url  [0]{\begingroup\@sanitize@url \@url }%
\providecommand \@url [1]{\endgroup\@href {#1}{\urlprefix }}%
\providecommand \urlprefix  [0]{URL }%
\providecommand \Eprint [0]{\href }%
\providecommand \doibase [0]{http://dx.doi.org/}%
\providecommand \selectlanguage [0]{\@gobble}%
\providecommand \bibinfo  [0]{\@secondoftwo}%
\providecommand \bibfield  [0]{\@secondoftwo}%
\providecommand \translation [1]{[#1]}%
\providecommand \BibitemOpen [0]{}%
\providecommand \bibitemStop [0]{}%
\providecommand \bibitemNoStop [0]{.\EOS\space}%
\providecommand \EOS [0]{\spacefactor3000\relax}%
\providecommand \BibitemShut  [1]{\csname bibitem#1\endcsname}%
\let\auto@bib@innerbib\@empty
\bibitem [{\citenamefont {Jaeger}(2001)}]{Jaeger2001}%
  \BibitemOpen
  \bibfield  {author} {\bibinfo {author} {\bibfnamefont {H.}~\bibnamefont
  {Jaeger}},\ }\href@noop {} {\enquote {\bibinfo {title} {The ``echo state''
  approach to analysing and training recurrent neural networks},}\ } (\bibinfo
  {year} {2001}),\ \bibinfo {note} {{G}MD-Report 148, GMD-German National
  Research Institute for Computer Science}\BibitemShut {NoStop}%
\bibitem [{\citenamefont {Maass}\ \emph {et~al.}(2002)\citenamefont {Maass},
  \citenamefont {Natschl\"{a}ger},\ and\ \citenamefont {Markram}}]{Maass2002}%
  \BibitemOpen
  \bibfield  {author} {\bibinfo {author} {\bibfnamefont {W.}~\bibnamefont
  {Maass}}, \bibinfo {author} {\bibfnamefont {T.}~\bibnamefont
  {Natschl\"{a}ger}}, \ and\ \bibinfo {author} {\bibfnamefont {H.}~\bibnamefont
  {Markram}},\ }\href@noop {} {\bibfield  {journal} {\bibinfo  {journal}
  {Neural Comput.}\ }\textbf {\bibinfo {volume} {14}},\ \bibinfo {pages} {2531}
  (\bibinfo {year} {2002})}\BibitemShut {NoStop}%
\bibitem [{\citenamefont {Jaeger}\ and\ \citenamefont
  {Haas}(2004)}]{Jaeger2004}%
  \BibitemOpen
  \bibfield  {author} {\bibinfo {author} {\bibfnamefont {H.}~\bibnamefont
  {Jaeger}}\ and\ \bibinfo {author} {\bibfnamefont {H.}~\bibnamefont {Haas}},\
  }\href@noop {} {\bibfield  {journal} {\bibinfo  {journal} {Science}\ }\textbf
  {\bibinfo {volume} {304}},\ \bibinfo {pages} {78} (\bibinfo {year}
  {2004})}\BibitemShut {NoStop}%
\bibitem [{\citenamefont {Verstraeten}\ \emph {et~al.}(2007)\citenamefont
  {Verstraeten}, \citenamefont {Schrauwen},\ and\ \citenamefont
  {Stroobandt}}]{Verstraeten2007}%
  \BibitemOpen
  \bibfield  {author} {\bibinfo {author} {\bibfnamefont {D.}~\bibnamefont
  {Verstraeten}}, \bibinfo {author} {\bibfnamefont {M.}~\bibnamefont
  {Schrauwen}, \bibfnamefont {B.~D'Haene}}, \ and\ \bibinfo {author}
  {\bibfnamefont {D.}~\bibnamefont {Stroobandt}},\ }\href@noop {} {\bibfield
  {journal} {\bibinfo  {journal} {Neural Netw.}\ }\textbf {\bibinfo {volume}
  {20}},\ \bibinfo {pages} {391} (\bibinfo {year} {2007})}\BibitemShut
  {NoStop}%
\bibitem [{\citenamefont {Luko\v{s}evi\v{c}ius}\ and\ \citenamefont
  {Jaeger}(2009)}]{Lukosevicius2009}%
  \BibitemOpen
  \bibfield  {author} {\bibinfo {author} {\bibfnamefont {M.}~\bibnamefont
  {Luko\v{s}evi\v{c}ius}}\ and\ \bibinfo {author} {\bibfnamefont
  {H.}~\bibnamefont {Jaeger}},\ }\href@noop {} {\bibfield  {journal} {\bibinfo
  {journal} {Comput. Sci. Rev.}\ }\textbf {\bibinfo {volume} {3}},\ \bibinfo
  {pages} {127} (\bibinfo {year} {2009})}\BibitemShut {NoStop}%
\bibitem [{\citenamefont {Pathak}\ \emph {et~al.}(2018)\citenamefont {Pathak},
  \citenamefont {Hunt}, \citenamefont {Girvan}, \citenamefont {Lu},\ and\
  \citenamefont {Ott}}]{Pathak2018}%
  \BibitemOpen
  \bibfield  {author} {\bibinfo {author} {\bibfnamefont {J.}~\bibnamefont
  {Pathak}}, \bibinfo {author} {\bibfnamefont {B.}~\bibnamefont {Hunt}},
  \bibinfo {author} {\bibfnamefont {M.}~\bibnamefont {Girvan}}, \bibinfo
  {author} {\bibfnamefont {Z.}~\bibnamefont {Lu}}, \ and\ \bibinfo {author}
  {\bibfnamefont {E.}~\bibnamefont {Ott}},\ }\href@noop {} {\bibfield
  {journal} {\bibinfo  {journal} {Phys. Rev. Lett.}\ }\textbf {\bibinfo
  {volume} {120}},\ \bibinfo {pages} {024102} (\bibinfo {year}
  {2018})}\BibitemShut {NoStop}%
\bibitem [{\citenamefont {Larger}\ \emph {et~al.}(2012)\citenamefont {Larger},
  \citenamefont {Soriano}, \citenamefont {Brunner}, \citenamefont {Appeltant},
  \citenamefont {Gutierrez}, \citenamefont {Pesquera}, \citenamefont
  {Mirasso},\ and\ \citenamefont {Fischer}}]{Larger2012}%
  \BibitemOpen
  \bibfield  {author} {\bibinfo {author} {\bibfnamefont {L.}~\bibnamefont
  {Larger}}, \bibinfo {author} {\bibfnamefont {M.~C.}\ \bibnamefont {Soriano}},
  \bibinfo {author} {\bibfnamefont {D.}~\bibnamefont {Brunner}}, \bibinfo
  {author} {\bibfnamefont {L.}~\bibnamefont {Appeltant}}, \bibinfo {author}
  {\bibfnamefont {J.~M.}\ \bibnamefont {Gutierrez}}, \bibinfo {author}
  {\bibfnamefont {L.}~\bibnamefont {Pesquera}}, \bibinfo {author}
  {\bibfnamefont {C.~R.}\ \bibnamefont {Mirasso}}, \ and\ \bibinfo {author}
  {\bibfnamefont {I.}~\bibnamefont {Fischer}},\ }\href@noop {} {\bibfield
  {journal} {\bibinfo  {journal} {Optics Express}\ }\textbf {\bibinfo {volume}
  {20}},\ \bibinfo {pages} {3241} (\bibinfo {year} {2012})}\BibitemShut
  {NoStop}%
\bibitem [{\citenamefont {Larger}\ \emph {et~al.}(2017)\citenamefont {Larger},
  \citenamefont {Bayl\'{o}n-Fuentes}, \citenamefont {Martinenghi},
  \citenamefont {Udaltsov}, \citenamefont {Chembo},\ and\ \citenamefont
  {Jacquot}}]{Larger2017}%
  \BibitemOpen
  \bibfield  {author} {\bibinfo {author} {\bibfnamefont {L.}~\bibnamefont
  {Larger}}, \bibinfo {author} {\bibfnamefont {A.}~\bibnamefont
  {Bayl\'{o}n-Fuentes}}, \bibinfo {author} {\bibfnamefont {R.}~\bibnamefont
  {Martinenghi}}, \bibinfo {author} {\bibfnamefont {V.~S.}\ \bibnamefont
  {Udaltsov}}, \bibinfo {author} {\bibfnamefont {Y.~K.}\ \bibnamefont
  {Chembo}}, \ and\ \bibinfo {author} {\bibfnamefont {M.}~\bibnamefont
  {Jacquot}},\ }\href@noop {} {\bibfield  {journal} {\bibinfo  {journal} {Phys.
  Rev. X}\ }\textbf {\bibinfo {volume} {7}},\ \bibinfo {pages} {011015}
  (\bibinfo {year} {2017})}\BibitemShut {NoStop}%
\bibitem [{\citenamefont {Torrejon}\ \emph {et~al.}(2017)\citenamefont
  {Torrejon}, \citenamefont {Riou}, \citenamefont {Araujo}, \citenamefont
  {Tsunegi}, \citenamefont {Khalsa}, \citenamefont {Querlioz}, \citenamefont
  {Bortolotti}, \citenamefont {Cros}, \citenamefont {Yakushiji}, \citenamefont
  {Fukushima}, \citenamefont {Kubota}, \citenamefont {Yuasa}, \citenamefont
  {Stiles},\ and\ \citenamefont {Grollier}}]{Torrejon2017}%
  \BibitemOpen
  \bibfield  {author} {\bibinfo {author} {\bibfnamefont {J.}~\bibnamefont
  {Torrejon}}, \bibinfo {author} {\bibfnamefont {M.}~\bibnamefont {Riou}},
  \bibinfo {author} {\bibfnamefont {F.~A.}\ \bibnamefont {Araujo}}, \bibinfo
  {author} {\bibfnamefont {S.}~\bibnamefont {Tsunegi}}, \bibinfo {author}
  {\bibfnamefont {G.}~\bibnamefont {Khalsa}}, \bibinfo {author} {\bibfnamefont
  {D.}~\bibnamefont {Querlioz}}, \bibinfo {author} {\bibfnamefont
  {P.}~\bibnamefont {Bortolotti}}, \bibinfo {author} {\bibfnamefont
  {V.}~\bibnamefont {Cros}}, \bibinfo {author} {\bibfnamefont {K.}~\bibnamefont
  {Yakushiji}}, \bibinfo {author} {\bibfnamefont {A.}~\bibnamefont
  {Fukushima}}, \bibinfo {author} {\bibfnamefont {H.}~\bibnamefont {Kubota}},
  \bibinfo {author} {\bibfnamefont {S.}~\bibnamefont {Yuasa}}, \bibinfo
  {author} {\bibfnamefont {M.~D.}\ \bibnamefont {Stiles}}, \ and\ \bibinfo
  {author} {\bibfnamefont {J.}~\bibnamefont {Grollier}},\ }\href@noop {}
  {\bibfield  {journal} {\bibinfo  {journal} {Nature}\ }\textbf {\bibinfo
  {volume} {547}},\ \bibinfo {pages} {428} (\bibinfo {year}
  {2017})}\BibitemShut {NoStop}%
\bibitem [{\citenamefont {Tsunegi}\ \emph {et~al.}(2019)\citenamefont
  {Tsunegi}, \citenamefont {Taniguchi}, \citenamefont {Nakajima}, \citenamefont
  {Miwa}, \citenamefont {Yakushiji}, \citenamefont {Fukushima}, \citenamefont
  {Yuasa},\ and\ \citenamefont {Kubota}}]{Tsunegi2019}%
  \BibitemOpen
  \bibfield  {author} {\bibinfo {author} {\bibfnamefont {S.}~\bibnamefont
  {Tsunegi}}, \bibinfo {author} {\bibfnamefont {T.}~\bibnamefont {Taniguchi}},
  \bibinfo {author} {\bibfnamefont {K.}~\bibnamefont {Nakajima}}, \bibinfo
  {author} {\bibfnamefont {S.}~\bibnamefont {Miwa}}, \bibinfo {author}
  {\bibfnamefont {K.}~\bibnamefont {Yakushiji}}, \bibinfo {author}
  {\bibfnamefont {A.}~\bibnamefont {Fukushima}}, \bibinfo {author}
  {\bibfnamefont {S.}~\bibnamefont {Yuasa}}, \ and\ \bibinfo {author}
  {\bibfnamefont {H.}~\bibnamefont {Kubota}},\ }\href@noop {} {\bibfield
  {journal} {\bibinfo  {journal} {Appl. Phys. Lett.}\ }\textbf {\bibinfo
  {volume} {114}},\ \bibinfo {pages} {164101} (\bibinfo {year}
  {2019})}\BibitemShut {NoStop}%
\bibitem [{\citenamefont {Nakajima}\ \emph {et~al.}(2018)\citenamefont
  {Nakajima}, \citenamefont {Hauser}, \citenamefont {Li},\ and\ \citenamefont
  {Pfeifer}}]{Nakajima2018}%
  \BibitemOpen
  \bibfield  {author} {\bibinfo {author} {\bibfnamefont {K.}~\bibnamefont
  {Nakajima}}, \bibinfo {author} {\bibfnamefont {H.}~\bibnamefont {Hauser}},
  \bibinfo {author} {\bibfnamefont {T.}~\bibnamefont {Li}}, \ and\ \bibinfo
  {author} {\bibfnamefont {R.}~\bibnamefont {Pfeifer}},\ }\href@noop {}
  {\bibfield  {journal} {\bibinfo  {journal} {Soft Robotics}\ }\textbf
  {\bibinfo {volume} {5}},\ \bibinfo {pages} {339} (\bibinfo {year}
  {2018})}\BibitemShut {NoStop}%
\bibitem [{\citenamefont {Fujii}\ and\ \citenamefont
  {Nakajima}(2017)}]{Fujii2017}%
  \BibitemOpen
  \bibfield  {author} {\bibinfo {author} {\bibfnamefont {K.}~\bibnamefont
  {Fujii}}\ and\ \bibinfo {author} {\bibfnamefont {K.}~\bibnamefont
  {Nakajima}},\ }\href@noop {} {\bibfield  {journal} {\bibinfo  {journal}
  {Phys. Rev. Appl.}\ }\textbf {\bibinfo {volume} {8}},\ \bibinfo {pages}
  {024030} (\bibinfo {year} {2017})}\BibitemShut {NoStop}%
\bibitem [{\citenamefont {Jaeger}(2002)}]{Jaeger2002}%
  \BibitemOpen
  \bibfield  {author} {\bibinfo {author} {\bibfnamefont {H.}~\bibnamefont
  {Jaeger}},\ }\href@noop {} {\enquote {\bibinfo {title} {Short term memory in
  echo state networks},}\ } (\bibinfo {year} {2002}),\ \bibinfo {note}
  {{G}MD-Report 152, GMD-German National Research Institute for Computer
  Science}\BibitemShut {NoStop}%
\bibitem [{\citenamefont {Dambre}\ \emph {et~al.}(2012)\citenamefont {Dambre},
  \citenamefont {Verstraeten}, \citenamefont {Schrauwen},\ and\ \citenamefont
  {Massar}}]{Dambre2012}%
  \BibitemOpen
  \bibfield  {author} {\bibinfo {author} {\bibfnamefont {J.}~\bibnamefont
  {Dambre}}, \bibinfo {author} {\bibfnamefont {D.}~\bibnamefont {Verstraeten}},
  \bibinfo {author} {\bibfnamefont {B.}~\bibnamefont {Schrauwen}}, \ and\
  \bibinfo {author} {\bibfnamefont {S.}~\bibnamefont {Massar}},\ }\href@noop {}
  {\bibfield  {journal} {\bibinfo  {journal} {Sci. Rep.}\ }\textbf {\bibinfo
  {volume} {2}},\ \bibinfo {pages} {514} (\bibinfo {year} {2012})}\BibitemShut
  {NoStop}%
\bibitem [{\citenamefont {Bertschinger}\ and\ \citenamefont
  {Natschl\"{a}ger}(2004)}]{Bertschinger2004}%
  \BibitemOpen
  \bibfield  {author} {\bibinfo {author} {\bibfnamefont {N.}~\bibnamefont
  {Bertschinger}}\ and\ \bibinfo {author} {\bibfnamefont {T.}~\bibnamefont
  {Natschl\"{a}ger}},\ }\href@noop {} {\bibfield  {journal} {\bibinfo
  {journal} {Neural Comput.}\ }\textbf {\bibinfo {volume} {16}},\ \bibinfo
  {pages} {1413} (\bibinfo {year} {2004})}\BibitemShut {NoStop}%
\bibitem [{\citenamefont {Boedecker}\ \emph {et~al.}(2012)\citenamefont
  {Boedecker}, \citenamefont {Obst}, \citenamefont {Lizier}, \citenamefont
  {Mayer},\ and\ \citenamefont {Asada}}]{Boedecker2012}%
  \BibitemOpen
  \bibfield  {author} {\bibinfo {author} {\bibfnamefont {J.}~\bibnamefont
  {Boedecker}}, \bibinfo {author} {\bibfnamefont {O.}~\bibnamefont {Obst}},
  \bibinfo {author} {\bibfnamefont {J.~T.}\ \bibnamefont {Lizier}}, \bibinfo
  {author} {\bibfnamefont {N.~M.}\ \bibnamefont {Mayer}}, \ and\ \bibinfo
  {author} {\bibfnamefont {M.}~\bibnamefont {Asada}},\ }\href@noop {}
  {\bibfield  {journal} {\bibinfo  {journal} {Theory Biosci.}\ }\textbf
  {\bibinfo {volume} {131}},\ \bibinfo {pages} {205} (\bibinfo {year}
  {2012})}\BibitemShut {NoStop}%
\bibitem [{\citenamefont {Farka\v{s}}\ \emph {et~al.}(2016)\citenamefont
  {Farka\v{s}}, \citenamefont {Bos\'{a}k},\ and\ \citenamefont
  {Gerge\v{l}}}]{Farkas2016}%
  \BibitemOpen
  \bibfield  {author} {\bibinfo {author} {\bibfnamefont {I.}~\bibnamefont
  {Farka\v{s}}}, \bibinfo {author} {\bibfnamefont {R.}~\bibnamefont
  {Bos\'{a}k}}, \ and\ \bibinfo {author} {\bibfnamefont {P.}~\bibnamefont
  {Gerge\v{l}}},\ }\href@noop {} {\bibfield  {journal} {\bibinfo  {journal}
  {Neural Netw.}\ }\textbf {\bibinfo {volume} {83}},\ \bibinfo {pages} {109}
  (\bibinfo {year} {2016})}\BibitemShut {NoStop}%
\bibitem [{\citenamefont {White}\ \emph {et~al.}(2004)\citenamefont {White},
  \citenamefont {Lee},\ and\ \citenamefont {Sompolinsky}}]{White2004}%
  \BibitemOpen
  \bibfield  {author} {\bibinfo {author} {\bibfnamefont {O.~L.}\ \bibnamefont
  {White}}, \bibinfo {author} {\bibfnamefont {D.~D.}\ \bibnamefont {Lee}}, \
  and\ \bibinfo {author} {\bibfnamefont {H.}~\bibnamefont {Sompolinsky}},\
  }\href@noop {} {\bibfield  {journal} {\bibinfo  {journal} {Phys. Rev. Lett.}\
  }\textbf {\bibinfo {volume} {92}},\ \bibinfo {pages} {148102} (\bibinfo
  {year} {2004})}\BibitemShut {NoStop}%
\bibitem [{\citenamefont {Rodan}\ and\ \citenamefont {Tino}(2011)}]{Rodan2011}%
  \BibitemOpen
  \bibfield  {author} {\bibinfo {author} {\bibfnamefont {A.}~\bibnamefont
  {Rodan}}\ and\ \bibinfo {author} {\bibfnamefont {P.}~\bibnamefont {Tino}},\
  }\href@noop {} {\bibfield  {journal} {\bibinfo  {journal} {IEEE Trans Neural
  Netw.}\ }\textbf {\bibinfo {volume} {22}},\ \bibinfo {pages} {131} (\bibinfo
  {year} {2011})}\BibitemShut {NoStop}%
\bibitem [{\citenamefont {Hermans}\ and\ \citenamefont
  {Schrauwen}(2010)}]{Hermans2010}%
  \BibitemOpen
  \bibfield  {author} {\bibinfo {author} {\bibfnamefont {M.}~\bibnamefont
  {Hermans}}\ and\ \bibinfo {author} {\bibfnamefont {B.}~\bibnamefont
  {Schrauwen}},\ }\href@noop {} {\bibfield  {journal} {\bibinfo  {journal}
  {Neural Netw.}\ }\textbf {\bibinfo {volume} {23}},\ \bibinfo {pages} {341}
  (\bibinfo {year} {2010})}\BibitemShut {NoStop}%
\bibitem [{\citenamefont {Marzen}(2017)}]{Marzen2017}%
  \BibitemOpen
  \bibfield  {author} {\bibinfo {author} {\bibfnamefont {S.}~\bibnamefont
  {Marzen}},\ }\href@noop {} {\bibfield  {journal} {\bibinfo  {journal} {Phys.
  Rev. E}\ }\textbf {\bibinfo {volume} {96}},\ \bibinfo {pages} {032308}
  (\bibinfo {year} {2017})}\BibitemShut {NoStop}%
\bibitem [{\citenamefont {Ganguli}\ \emph {et~al.}(2008)\citenamefont
  {Ganguli}, \citenamefont {Huh},\ and\ \citenamefont
  {Sompolinsky}}]{Ganguli2008}%
  \BibitemOpen
  \bibfield  {author} {\bibinfo {author} {\bibfnamefont {S.}~\bibnamefont
  {Ganguli}}, \bibinfo {author} {\bibfnamefont {D.}~\bibnamefont {Huh}}, \ and\
  \bibinfo {author} {\bibfnamefont {H.}~\bibnamefont {Sompolinsky}},\
  }\href@noop {} {\bibfield  {journal} {\bibinfo  {journal} {Proc. Natl. Acad.
  Sci. USA}\ }\textbf {\bibinfo {volume} {105}},\ \bibinfo {pages} {18970}
  (\bibinfo {year} {2008})}\BibitemShut {NoStop}%
\bibitem [{\citenamefont {Schuecker}\ \emph {et~al.}(2018)\citenamefont
  {Schuecker}, \citenamefont {Goedeke},\ and\ \citenamefont
  {Helias}}]{Schuecker2018}%
  \BibitemOpen
  \bibfield  {author} {\bibinfo {author} {\bibfnamefont {J.}~\bibnamefont
  {Schuecker}}, \bibinfo {author} {\bibfnamefont {S.}~\bibnamefont {Goedeke}},
  \ and\ \bibinfo {author} {\bibfnamefont {M.}~\bibnamefont {Helias}},\
  }\href@noop {} {\bibfield  {journal} {\bibinfo  {journal} {Phys. Rev. X}\
  }\textbf {\bibinfo {volume} {8}},\ \bibinfo {pages} {041029} (\bibinfo {year}
  {2018})}\BibitemShut {NoStop}%
\bibitem [{\citenamefont {Sompolinsky}\ \emph {et~al.}(1988)\citenamefont
  {Sompolinsky}, \citenamefont {Crisanti},\ and\ \citenamefont
  {Sommers}}]{Sompolinsky1988}%
  \BibitemOpen
  \bibfield  {author} {\bibinfo {author} {\bibfnamefont {H.}~\bibnamefont
  {Sompolinsky}}, \bibinfo {author} {\bibfnamefont {A.}~\bibnamefont
  {Crisanti}}, \ and\ \bibinfo {author} {\bibfnamefont {H.~J.}\ \bibnamefont
  {Sommers}},\ }\href@noop {} {\bibfield  {journal} {\bibinfo  {journal} {Phys.
  Rev. Lett.}\ }\textbf {\bibinfo {volume} {61}},\ \bibinfo {pages} {259}
  (\bibinfo {year} {1988})}\BibitemShut {NoStop}%
\bibitem [{\citenamefont {Toyoizumi}\ and\ \citenamefont
  {Abbott}(2011)}]{Toyoizumi2011}%
  \BibitemOpen
  \bibfield  {author} {\bibinfo {author} {\bibfnamefont {T.}~\bibnamefont
  {Toyoizumi}}\ and\ \bibinfo {author} {\bibfnamefont {L.~F.}\ \bibnamefont
  {Abbott}},\ }\href@noop {} {\bibfield  {journal} {\bibinfo  {journal} {Phys.
  Rev. E}\ }\textbf {\bibinfo {volume} {84}},\ \bibinfo {pages} {051908}
  (\bibinfo {year} {2011})}\BibitemShut {NoStop}%
\bibitem [{\citenamefont {Cessac}\ and\ \citenamefont
  {Samuelides}(2007)}]{Cessac2007}%
  \BibitemOpen
  \bibfield  {author} {\bibinfo {author} {\bibfnamefont {B.}~\bibnamefont
  {Cessac}}\ and\ \bibinfo {author} {\bibfnamefont {M.}~\bibnamefont
  {Samuelides}},\ }\href@noop {} {\bibfield  {journal} {\bibinfo  {journal}
  {Eur. Phys. J. Spec. Top.}\ }\textbf {\bibinfo {volume} {142}},\ \bibinfo
  {pages} {7} (\bibinfo {year} {2007})}\BibitemShut {NoStop}%
\bibitem [{\citenamefont {Massar}\ and\ \citenamefont
  {Massar}(2013)}]{Massar2013}%
  \BibitemOpen
  \bibfield  {author} {\bibinfo {author} {\bibfnamefont {M.}~\bibnamefont
  {Massar}}\ and\ \bibinfo {author} {\bibfnamefont {S.}~\bibnamefont
  {Massar}},\ }\href@noop {} {\bibfield  {journal} {\bibinfo  {journal} {Phys.
  Rev. E}\ }\textbf {\bibinfo {volume} {87}},\ \bibinfo {pages} {042809}
  (\bibinfo {year} {2013})}\BibitemShut {NoStop}%
\bibitem [{\citenamefont {Molgedey}\ \emph {et~al.}(1992)\citenamefont
  {Molgedey}, \citenamefont {Schuchhardt},\ and\ \citenamefont
  {Schuster}}]{Molgedey1992}%
  \BibitemOpen
  \bibfield  {author} {\bibinfo {author} {\bibfnamefont {L.}~\bibnamefont
  {Molgedey}}, \bibinfo {author} {\bibfnamefont {J.}~\bibnamefont
  {Schuchhardt}}, \ and\ \bibinfo {author} {\bibfnamefont {H.~G.}\ \bibnamefont
  {Schuster}},\ }\href@noop {} {\bibfield  {journal} {\bibinfo  {journal}
  {Phys. Rev. Lett.}\ }\textbf {\bibinfo {volume} {69}},\ \bibinfo {pages}
  {3717} (\bibinfo {year} {1992})}\BibitemShut {NoStop}%
\bibitem [{\citenamefont {Cover}\ and\ \citenamefont
  {Thomas}(2006)}]{Cover2006}%
  \BibitemOpen
  \bibfield  {author} {\bibinfo {author} {\bibfnamefont {T.~M.}\ \bibnamefont
  {Cover}}\ and\ \bibinfo {author} {\bibfnamefont {J.~A.}\ \bibnamefont
  {Thomas}},\ }\href@noop {} {\emph {\bibinfo {title} {Elements of Information
  Theory, 2nd ed.}}}\ (\bibinfo  {publisher} {John Wiley \& Sons},\ \bibinfo
  {address} {Hoboken, NJ},\ \bibinfo {year} {2006})\BibitemShut {NoStop}%
\bibitem [{\citenamefont {Prokopenko}\ \emph {et~al.}(2011)\citenamefont
  {Prokopenko}, \citenamefont {Lizier}, \citenamefont {Obst},\ and\
  \citenamefont {Wang}}]{Prokopenko2011}%
  \BibitemOpen
  \bibfield  {author} {\bibinfo {author} {\bibfnamefont {M.}~\bibnamefont
  {Prokopenko}}, \bibinfo {author} {\bibfnamefont {J.~T.}\ \bibnamefont
  {Lizier}}, \bibinfo {author} {\bibfnamefont {O.}~\bibnamefont {Obst}}, \ and\
  \bibinfo {author} {\bibfnamefont {X.~R.}\ \bibnamefont {Wang}},\ }\href@noop
  {} {\bibfield  {journal} {\bibinfo  {journal} {Phys. Rev. E}\ }\textbf
  {\bibinfo {volume} {84}},\ \bibinfo {pages} {041116} (\bibinfo {year}
  {2011})}\BibitemShut {NoStop}%
\bibitem [{\citenamefont {Livi}\ \emph {et~al.}(2017)\citenamefont {Livi},
  \citenamefont {Bianchi},\ and\ \citenamefont {Alippi}}]{Livi2017}%
  \BibitemOpen
  \bibfield  {author} {\bibinfo {author} {\bibfnamefont {L.}~\bibnamefont
  {Livi}}, \bibinfo {author} {\bibfnamefont {F.~M.}\ \bibnamefont {Bianchi}}, \
  and\ \bibinfo {author} {\bibfnamefont {C.}~\bibnamefont {Alippi}},\
  }\href@noop {} {\bibfield  {journal} {\bibinfo  {journal} {IEEE Trans. Neural
  Netw. Learn. Syst.}\ }\textbf {\bibinfo {volume} {29}},\ \bibinfo {pages}
  {706} (\bibinfo {year} {2017})}\BibitemShut {NoStop}%
\bibitem [{\citenamefont {Marquez}\ \emph {et~al.}(2018)\citenamefont
  {Marquez}, \citenamefont {Larger}, \citenamefont {Jacquot}, \citenamefont
  {Chembo},\ and\ \citenamefont {Brunner}}]{Marquez2018}%
  \BibitemOpen
  \bibfield  {author} {\bibinfo {author} {\bibfnamefont {B.~A.}\ \bibnamefont
  {Marquez}}, \bibinfo {author} {\bibfnamefont {L.}~\bibnamefont {Larger}},
  \bibinfo {author} {\bibfnamefont {M.}~\bibnamefont {Jacquot}}, \bibinfo
  {author} {\bibfnamefont {Y.~K.}\ \bibnamefont {Chembo}}, \ and\ \bibinfo
  {author} {\bibfnamefont {D.}~\bibnamefont {Brunner}},\ }\href@noop {}
  {\bibfield  {journal} {\bibinfo  {journal} {Sci. Rep.}\ }\textbf {\bibinfo
  {volume} {8}},\ \bibinfo {pages} {3319} (\bibinfo {year} {2018})}\BibitemShut
  {NoStop}%
\bibitem [{\citenamefont {Sussillo}\ and\ \citenamefont
  {Abbott}(2009)}]{Sussillo2009}%
  \BibitemOpen
  \bibfield  {author} {\bibinfo {author} {\bibfnamefont {D.}~\bibnamefont
  {Sussillo}}\ and\ \bibinfo {author} {\bibfnamefont {L.~F.}\ \bibnamefont
  {Abbott}},\ }\href@noop {} {\bibfield  {journal} {\bibinfo  {journal}
  {Neuron}\ }\textbf {\bibinfo {volume} {63}},\ \bibinfo {pages} {544}
  (\bibinfo {year} {2009})}\BibitemShut {NoStop}%
\bibitem [{\citenamefont {Jaeger}\ \emph {et~al.}(2007)\citenamefont {Jaeger},
  \citenamefont {Luko\v{s}evi\v{c}ius}, \citenamefont {Popovici},\ and\
  \citenamefont {Siewert}}]{Jaeger2007}%
  \BibitemOpen
  \bibfield  {author} {\bibinfo {author} {\bibfnamefont {H.}~\bibnamefont
  {Jaeger}}, \bibinfo {author} {\bibfnamefont {M.}~\bibnamefont
  {Luko\v{s}evi\v{c}ius}}, \bibinfo {author} {\bibfnamefont {D.}~\bibnamefont
  {Popovici}}, \ and\ \bibinfo {author} {\bibfnamefont {U.}~\bibnamefont
  {Siewert}},\ }\href@noop {} {\bibfield  {journal} {\bibinfo  {journal}
  {Neural Netw.}\ }\textbf {\bibinfo {volume} {20}},\ \bibinfo {pages} {335}
  (\bibinfo {year} {2007})}\BibitemShut {NoStop}%
\bibitem [{\citenamefont {Inubushi}\ and\ \citenamefont
  {Yoshimura}(2017)}]{Inubushi2017}%
  \BibitemOpen
  \bibfield  {author} {\bibinfo {author} {\bibfnamefont {M.}~\bibnamefont
  {Inubushi}}\ and\ \bibinfo {author} {\bibfnamefont {K.}~\bibnamefont
  {Yoshimura}},\ }\href@noop {} {\bibfield  {journal} {\bibinfo  {journal}
  {Sci. Rep.}\ }\textbf {\bibinfo {volume} {7}},\ \bibinfo {pages} {10199}
  (\bibinfo {year} {2017})}\BibitemShut {NoStop}%
\end{thebibliography}

%

\end{document}